\documentclass[12pt]{iopart}
\usepackage{graphicx}
\usepackage{iopams}
\usepackage{setstack}
\usepackage{amsfonts,amssymb,graphicx}
\usepackage{mathrsfs}
\usepackage{amsthm}
\usepackage[usenames]{color}

\begin{document}
	\newcommand{\eg}{\textit{e.g.}{~}}
	\newcommand{\vphi}{\varphi}
	\newcommand{\bq}{\begin{equation}}
	\newcommand{\ba}{\begin{eqnarray}}
	\newcommand{\eq}{\end{equation}}
	\newcommand{\ee}{\end{equation}}
\newcommand{\ea}{\end{eqnarray}}
\newcommand{\tchi} {{\tilde \chi}}
\newcommand{\tA} {{\tilde A}}
\newcommand{\sech} { {\rm sech}}
\newcommand{\pstar}{\mbox{$\psi^{\ast}$}}
\newcommand {\tu} {{\tilde u}}
\newcommand {\tv} {{\tilde v}}
\newcommand{\dq}{{\dot q}}

\title[]{Soliton dynamics in  the ABS nonlinear spinor model with external fields}

\author{Franz G.  Mertens$^1$ , Bernardo Sánchez-Rey$^2$ and Niurka R. Quintero$^2$} 
\address{$^1$ Physikalisches Institut, Universit\"at Bayreuth, D-95440 Bayreuth, Germany}
\address{$^2$ Department of Applied Physics I, 
EPS. University of Seville, 41011 Seville, Spain} 
\eads{\mailto{franzgmertens@gmail.com},\mailto{bernardo@us.es},\mailto{niurka@us.es}}

\vspace{10pt}
\begin{indented}
\item[]July 2021
\end{indented}

\begin{abstract}
We consider  the novel nonlinear model in ($1+1$)-dimensions for Dirac spinors recently introduced by Alexeeva, Barashenkov, and Saxena \cite{alexeeva:2019} (ABS model), which admits an exact explicit solitary-wave (soliton for short) solution. The charge, the momentum, and the energy of this solution are conserved. We investigate the dynamics of the soliton subjected to several potentials: a ramp, a harmonic, and a periodic potential. We develop a Collective Coordinates Theory by making an ansatz for a moving soliton where the position, rapidity, and momentum, are functions of time. We insert the ansatz into the Lagrangian density of the model, integrate over space and obtain a Lagrangian as a function of the collective coordinates.  This Lagrangian differs only in the charge and mass with the Lagrangian of a collective coordinates theory for the  Gross-Neveu equation. Thus the soliton dynamics in the  ABS spinor model is \textit{qualitatively} the same as in the  Gross-Neveu equation, but 
\textit{quantitatively} it differs. These results of the collective coordinates theory are  confirmed by simulations, i.e., by numerical solutions for solitons of the   ABS spinor model, subjected to the above potentials.
\end{abstract}

\submitto{\JPA}

\vspace{2pc}
\noindent{\it Keywords}: solitons, collective coordinates, nonlinear Dirac equation, potentials

\maketitle

\section{\label{sec1}Introduction}

In the theory of solitons (or solitary waves), nonlinearity plays a crucial role. It compensates the dispersive effects so that the soliton emerges and travels with constant velocity and without changing
its shape \cite{drazin:1989,scott:1999,dauxois:2006}. The existence and stability of  solitons
is a very important issue, for example for all-optical high-speed transmission systems, where it is necessary to
provide a stable information transfer \cite{hasegawa:1997,hasegawa:2003}.
A tiny change in the nonlinear terms can produce not only the loss of integrability
\cite{remoissenet:1981,peyrard:1982}, but also the instability of the system \cite{vakhitov:1973,shao:2014} with a subsequent loss of information. Additionally, by changing nonlinear terms new phenomena, such as the well known soliton ratchets \cite{salerno:2002} and the Feschbach resonance \cite{kevrekidis:2003}, can arise.

It is also well stablished that solitons can persist under certain spatio-temporal perturbations. In some cases, even exact solutions have been found  \cite{chen:1976,dominguez:1993,serkin:2007}. However, in general, only approximate solutions are found by using variational methods, the method of moments, or the generalized traveling-wave method \cite{quintero:2010}. In particular, soliton dynamics under  external fields and dissipation has been extensively studied for the case of topological solitons (sine-Gordon, $\phi^4$, double sine-Gordon), non-topological solitons (Heisenberg model) \cite{mertens:1997}, and envelope solitons (nonlinear Schr\"odinger equation)  \cite{fogel:1977,rice:1980,salerno:1982,kivshar:1989,quintero:2000}. The collective coordinates approach usually maps the evolution of these extended systems into the evolution of the soliton center of mass. But in the case of the  Gross-Neveu soliton under external fields, the collective coordinates theory becomes more complex due to the \textit{larger} number of 
collective coordinates required to reach a satisfactory approximation   \cite{mertens:2012,quintero:2019,quintero:2019a}. 

 Soliton solutions for different types of nonlinearity, their stability, and  conservation laws, have been studied in  \cite{esteban:1995,pelinovsky:2014,cuevas:2016a,sabirov:2018,borrelli:2021}. Very recently, Alexeeva, Barashenkov, and Saxena \cite{alexeeva:2019}  studied various $\mathcal{PT}$ symmetric spinor models belonging to the general family of the nonlinear Dirac equation. In particular, they 
introduced a new nonlinear model in ($1+1$)-dimensions for the Dirac spinor
\begin{equation} \label{eq1}
\Psi(x,t)=
\left(\!
\begin{array}{c}
u(x,t) \\ 
v(x,t) 
\end{array}
\!\right),
\end{equation}
which admits an exact explicit solitary-wave solution. The components $u$ and $v$ fulfill the nonlinear PDEs
\begin{eqnarray}
\label{eq2a}
i\,(u_t-u_x)+v+u^\star\,v^2&=&0, \\
\label{eq2b}
i\,(v_t+v_x)+u+v^\star\,u^2&=&0. 
\end{eqnarray} 
When written in covariant  notation, this reads
\begin{eqnarray}
\label{eq4}
i\,\gamma^{\mu}\,\partial_{\mu}\,\Psi+\Psi+(\bar{\Psi}\,\Psi)\,\Psi-\frac{1}{2}\,J_{\mu}\,\gamma^{\mu} \Psi&=&0. 
\end{eqnarray}
Hereinafter, we refer to  Eqs.\ (\ref{eq2a})-(\ref{eq2b}) or equivalently Eq.\ (\ref{eq4})
as the Alexeeva-Barashenkov-Saxena (ABS) model.  
Here, $\bar{\Psi}=\Psi^\dagger  \gamma^0$ is the conjugate spinor,  $\gamma^{\mu}$  are the Dirac gamma matrices for which the following representation is used
\begin{eqnarray} \label{q1}
\gamma^{0} = 
\left( \begin{array}{cc}
0 & 1  \\
1 & 0  \end{array} \right), \qquad \gamma^{1} = 
\left( \begin{array}{cc}
0 & 1  \\
-1 & 0  \end{array} \right)\,,
\end{eqnarray}
and  $J_{\mu}=\bar{\Psi}\,\gamma_{\mu}\,\Psi$, with $\gamma_0=\gamma^{0}$ and $\gamma_1=-\gamma^{1}$. 

The linear terms in Eqs.\ (\ref{eq2a}) and (\ref{eq2b}) are the same as the linear terms in the single-component massive Gross-Neveu model:
\begin{eqnarray} \label{eq6new}
i\,(u_t-u_x)+v+(u\,v^\star+u^\star\,v)\,v&=&0, \\
\label{eq7new}
i\,(v_t+v_x)+u+(v\,u^\star+v^\star\,u)\,u&=&0.
\end{eqnarray}
In covariant notation this is 
\begin{equation}
\label{eqnew8}
i\,\gamma^\mu\,\partial_{\mu} \Psi+\Psi+(\bar{\Psi}\,\Psi)\,\Psi=0.
\end{equation}
This means that only the nonlinear terms in the  ABS spinor model differ from the nonlinear terms in the   Gross-Neveu equation. 

The  Gross-Neveu equation has an exact explicit solitary-wave solution. Its dynamics in an external field was studied by several authors 
 \cite{alvarez:1981,dominguez:1993,nogami:1995,mertens:2012}. In particular, the motion of the soliton in several simple potentials (ramp, harmonic, and periodic potential) was investigated \cite{nogami:1995,mertens:2012} by using a Collective Coordinates (CC) Theory. The results were confirmed by a very good agreement with simulations (i.e., numerical solutions of the  Gross-Neveu equation with a potential, using the exact soliton solution as initial condition). 

The goal of our paper is to investigate the dynamics of a soliton in the  ABS nonlinear spinor model, Eqs. (\ref{eq2a}) and (\ref{eq2b}), subjected to the same potentials as  laid out above. 
We will show that the CC Theory of the  ABS spinor model yields a CC Lagrangian that (apart from the coefficients and an irrelevant phase) has the same expression as the CC Lagrangian of the  Gross-Neveu equation in Ref. \cite{mertens:2012}. This means that the dynamics of the  ABS spinor soliton is \textit{qualitatively} the same as the dynamics of the  Gross-Neveu soliton. \textit{Quantitatively} the dynamics of the two models differ due to the difference in the coefficients in the two Lagrangians, since they are defined as integrals over expressions which contain the soliton solutions. This result is  worthy of note, because the dynamics of solitons in models with different nonlinearities are  normally \textit{qualitatively} different. A good example of this fact is  given in the nonlinear Schr\"odinger (NLS) equation \cite{cooper:2012,dawson:2017}. Moreover, note that the spectrum of the Hamiltonian of the NLD
systems extends indefinitely in both directions, positive and
negative \cite{barashenkov:1983,barashenkov:1998,pelinovsky:2014}.
However, the variational approach used here does not assume the
existence of any minima. It is equivalent to a projection technique
rather than minimizing the Hamiltonian (see the appendix
\ref{apA}).

Our paper is organized as follows: in Section \ref{sec2}, the conservation of the charge, the momentum, and the energy of the   ABS nonlinear spinor model are derived. The stationary soliton solution found in \cite{alexeeva:2019} is then presented and a Lorentz boost yields a moving soliton. In Section \ref{sec3}, electromagnetic interactions are introduced through a gauge covariant derivative. The axial gauge where the zero component is the potential $V(x)$  is chosen, and a new Lagrangian together with new conservation laws are obtained. An ansatz for a trial wave function is introduced in Section \ref{sec4}. The ansatz is made on the assumption that it has the same functional form as the exact moving solution of the unperturbed system, and depends on time only through the CCs describing the position, rapidity, and momentum of the soliton. This trial wave function is inserted into the Lagrangian density
of the model. Integration over space yields a CC Lagrangian which   differs only in the coefficients with respect to the CC Lagrangian of the Gross-Neveu equation in Ref. \cite{mertens:2012}. The three Lagrange equations finally result in a Newtonian-like equation of motion for the soliton center of mass. Moreover the soliton charge, momentum, and energy are calculated. Finally, in Section  \ref{sec5}, the three  Lagrange equations are solved analytically or numerically for the cases of a ramp, a harmonic, and a periodic potential $V(x)$. The results are compared with simulations, that is, numerical solutions of  the ABS nonlinear spinor equations with a potential $V(x)$, and a very good agreement is found. 

\section{ Lagrangian density, conservation laws and solitary-wave solution of the ABS spinor model. \label{sec2}}

The nonlinear Eqs.\ (\ref{eq2a}) and (\ref{eq2b}) for the Dirac spinor can be derived from the Lagrangian density 
 \begin{eqnarray}
\mathcal{L} &=& \frac{i}{2}\,\left[\bar{\Psi}\,\gamma^{\mu}\,\partial_{\mu}\,\Psi-\partial_{\mu}\,
\bar{\Psi}\,\gamma^{\mu}\,\Psi\right]+\bar{\Psi}\,\Psi+\frac{1}{2}\,(\bar{\Psi}\,\Psi)^2-\frac{1}{4}\,J_{\mu}\,J^{\mu} \label{eqL}\\
&=&
 \left(\frac{i}{2}\right) [(u_t-u_x)\,u^\star -(u_t^\star-u_x^\star)\,u+
(v_t+v_x)\,v^\star -(v_t^\star+v_x^\star)\,v] + \nonumber \\
&+& u\,v^\star + u^\star\,v+\frac{1}{2} \left[(u\,v^\star)^2+(u^\star\,v)^2\right]. 
\label{eq5}
\end{eqnarray}
 Indeed, the corresponding Euler-Lagrange equations 
\begin{eqnarray} \label{eq4a}
\frac{d\,}{dt}\,\frac{\partial {\mathcal L} \quad}{\partial u_t^\star}+
\frac{d\,}{dx}\,\frac{\partial {\mathcal L} \quad}{\partial u_x^\star}-
\frac{\partial {\mathcal L}}{\partial u^\star}&=& 0, \\
\label{eq4b}
\frac{d\,}{dt}\,\frac{\partial {\mathcal L} \quad}{\partial v_t^\star}+
\frac{d\,}{dx}\,\frac{\partial {\mathcal L} \quad}{\partial v_x^\star}-
\frac{\partial {\mathcal L}}{\partial v^\star}&=& 0,
\end{eqnarray}
yield Eq. (\ref{eq2a}) and Eq. (\ref{eq2b}), respectively. 

A first conservation law can be derived by the following steps:  \textit{i)} multiply Eq.\ (\ref{eq2a}) by $u^\star$, and substract the complex conjugate of Eq. (\ref{eq2a}) multiplied by $u$; \textit{ii)} apply the same procedure to Eq. (\ref{eq2b}) multiplying by $v^\star$ 
and $v$, respectively;  \textit{iii)} add  the results of the previous two steps. The continuity equation
\begin{equation}
\label{eq5a}
\rho_t+j_x=0,
\end{equation}
is then obtained, where 
\begin{equation}
\label{eq5b}
\rho(x,t)=|u|^2+|v|^2,
\end{equation}  
is the charge density and 
\begin{equation}
\label{eq5c}
j(x,t)=|v|^2-|u|^2,
\end{equation}
is the current density. The charge 
\begin{equation}
\label{eq5d}
Q=\int_{-\infty}^{+\infty}\,dx\,\rho(x,t),
\end{equation}
is conserved, assuming $j(+\infty,t)-j(-\infty,t)=0$. 

In a similar way, namely by multiplying Eqs.\ (\ref{eq2a}) and (\ref{eq2b}) by the space derivative of the fields, and adding in step \textit{i)} instead of substracting, we derive the conservation law for the momentum
\begin{equation}
\label{eq6a}
\mathcal{P}_t+\Phi_x=0, 
\end{equation}
 where
\begin{eqnarray}
\label{eq6b} 
\mathcal{P}(x,t)&=&\frac{i}{2}\,[u_x^\star\,u-u_x\,u^\star+v_x^\star\,v-v_x\,v^\star] ,\\
\label{eq6c}
\Phi(x,t)&=&\frac{-i}{2}\,[u_t^\star\,u-u_t\,u^\star+v_t^\star\,v-v_t\,v^\star]+\\ \nonumber
&& (u\,v^\star+u^\star\,v)+\frac{1}{2}[(u\,v^\star)^2+(u^\star\,v)^2], 
\end{eqnarray}
 are, respectively, the momentum density and the momentum current. Clearly the momentum 
\begin{equation}
	\label{eqP}
	P=\int_{-\infty}^{+\infty}\,dx\,\mathcal{P}(x,t),
\end{equation} 	
will be a conserved quantity if $\Phi(+\infty,t)-\Phi(-\infty,t)=0$.	
 	
A third conservation law for the energy can be obtained with the same procedure, but using the time derivative of the fields: 
\begin{equation}
\label{eq7a}
\mathcal{H}_t+\mathcal{J}_x=0,
\end{equation}
 where now 
\begin{eqnarray} 
\label{eq7b} 
\mathcal{H}(x,t)&=&\frac{i}{2}\,[u^\star\,u_x-u\,u_x^\star-v^\star\,v_x+v\,v_x^\star]-\\ \nonumber
&&
(u\,v^\star+u^\star\,v)-\frac{1}{2}[(u\,v^\star)^2+(u^\star\,v)^2] 
\end{eqnarray}
 is the energy density, and
\begin{equation}
\label{eq7c}
\mathcal{J}(x,t)=\frac{i}{2}\,[u_t^\star\,u-u_t\,u^\star+v_t\,v^\star-v_t^\star\,v]
\end{equation}
the  energy current density. The energy 
\begin{equation}
	\label{eqE}
	E=\int_{-\infty}^{+\infty}\,dx\,\mathcal{H}(x,t)
\end{equation} 
will be conserved if $\mathcal{J}(+\infty,t)-\mathcal{J}(-\infty,t)=0$.

 At this point, we introduce the stationary soliton solution derived in Ref. \cite{alexeeva:2019}:
\begin{eqnarray}
\label{eq8a}
u(x,t)&=& a(x)\,\e^{i\,\theta(x)}\,\e^{-i\,\omega\,t}\\
\label{eq8b}
v(x,t)&=& -a(x)\,\e^{-i\,\theta(x)}\,\e^{-i\,\omega\,t}
\end{eqnarray}
where
\begin{equation}\label{eq9}
a^2=2\,\frac{\cos(2\,\theta)-\omega}{\cos(4\,\theta)}=-\frac{2\,\theta_x}{\cos(4\,\theta)}, 
\end{equation}
with 
\begin{eqnarray}\label{eq10a}
\theta(x)&=&-\arctan[\lambda\,\tanh(\kappa\,x)], \\
\lambda&=&\sqrt{\frac{1-\omega}{1+\omega}}, \qquad \kappa=\sqrt{1-\omega^2}.
\end{eqnarray}
Note that $\lambda$ and $\kappa$ depend on the frequency $\omega$, which is the only parameter of the solution. The charge density is
\begin{equation}\label{eq11}
\rho(x)=2 a^2=4 (1-\omega) \sech^2(\kappa x) \,
\frac{1+\lambda^2\,\tanh^2(\kappa\,x)}{1-6 \lambda^2\,\tanh^2(\kappa\,x)+\lambda^4\,\tanh^4(\kappa x)},
\end{equation}
and the charge reads 
\begin{equation}\label{eq12}
Q=2\,\ln\frac{\omega+\kappa}{\omega-\kappa}.
\end{equation}
The momentum of the stationary soliton is zero, and for its energy we obtain 
	\begin{equation}\label{eqE0}
		E=\sqrt{2}\,\ln\frac{1+\sqrt{2}\,\kappa}{1-\sqrt{2}\,\kappa}.
	\end{equation}

The range of admissible frequencies is
\begin{equation}\label{eq13}
\frac{1}{\sqrt{2}}\,<\,\omega\,<\, 1.  
\end{equation}
The charge density is bell-shaped for $\frac{3}{4}\,\le\,\omega\,<\, 1$, and has two humps  for $\frac{1}{\sqrt{2}}\,<\,\omega\,<\frac{3}{4}$   at $x=\pm x_{m}$ given by
\begin{equation}\label{eq14}
\tanh^2(\kappa\,x_{m})=\frac{1+\omega}{1-\omega}\,\frac{1-\omega-\sqrt{\omega^2-1/2}}
{1+\omega+\sqrt{\omega^2-1/2}}.
\end{equation}

 A moving soliton can be obtained by using the Lorentz transformation  
 \begin{eqnarray}
 \label{eq15a}
x' &=&\gamma\,(x-v_s\,t)=\cosh(\beta)\,\cdot\,x-\sinh(\beta)\,\cdot\,t, \\ t'&=&\gamma\,(t-v_s\,x)=\cosh(\beta)\,\cdot\,t-\sinh(\beta)\,\cdot\,x,
\label{eq15b}
\end{eqnarray} 
where
\begin{equation}\label{eq16}
\gamma=\frac{1}{\sqrt{1-v_s^2}}=\cosh(\beta)
\end{equation}
with the soliton velocity $v_s$ and the rapidity $\beta$. Thus the transformation matrix for the space-time vector is 
\begin{equation} \label{eq17}
\Lambda=
\left(\!
\begin{array}{cc}
\cosh(\beta) \quad & -\sinh(\beta) \\ 
-\sinh(\beta) \quad &\cosh(\beta)
\end{array}
\!\right).
\end{equation}
The transformation of the Dirac spinor $\Psi(x,t)$ is  
\begin{equation}\label{eq17a}
\Psi(x,t)=S\,\Psi'(x',t')
\end{equation}
where the matrix $S$ satisfies
\begin{equation}\label{eq18}
S\,\gamma^{\mu}\,S^{-1}=\Lambda^{\mu}_{\,\rho}\,\gamma^{\rho}.
\end{equation}
Using the representation (\ref{eq4}) for the $\gamma$-matrices, we obtain 
\begin{equation} \label{eq19}
S=
\left(\!
\begin{array}{cc}
\e^{-\beta/2} \quad & 0 \\ 
0 \quad & \e^{\beta/2}
\end{array}
\!\right).
\end{equation}
Thus the moving-soliton solution is  
\begin{eqnarray} \label{eq20a}
u(x,t) &=& \e^{-\beta/2} \,a(x')\,\e^{i\,\theta(x')}\,\e^{-i\,\omega\,t'},\\ 
\label{eq20b}
v(x,t) &=& -\e^{\beta/2} \,a(x')\,\e^{-i\,\theta(x')}\,\e^{-i\,\omega\,t'},
\end{eqnarray}
with $x',t'$ from Eqs. (\ref{eq15a})-(\ref{eq15b}).  Notice that the width of the charge density  
\begin{equation}
\rho(x,t)=2\,\gamma\,a^2[\gamma\,(x-v_s\,t)]   
\label{eqrho}
\end{equation} 
is Lorentz-contracted, while the height is increased by the factor $\gamma$ such that the charge is conserved. 

\section{ ABS spinor model with external fields} \label{sec3}

Electromagnetic interactions can be introduced through the gauge covariant derivative
\begin{equation} \label{eq21}
i\,\partial_{\mu}\,\Psi \to (i\,\partial_{\mu}-e\,A_{\mu})\,\Psi.
\end{equation}
Consequently, the Lagragian density contains the additional term  
\begin{equation}
	\mathcal{L}_{3} = -e\,\bar{\Psi}\,\gamma^{\mu}\,A_{\mu}\,\Psi=
	-e\,A_0\,\bar{\Psi}\,\gamma^{0}\,\Psi-e\,A_{1}\,\bar{\Psi}\,\gamma^{1}\,\Psi. 
	\label{eq23}
\end{equation}
The new Lagrangian
\begin{equation}
	\tilde\mathcal{{L}} = \mathcal{L}+\mathcal{L}_{3},
	\label{eq23a}
\end{equation}
where $\mathcal{L}$ is the unperturbed Lagrangian density given by Eq.\,(\ref{eqL}), is invariant under the combined transformations 
\begin{equation}
\Psi \to \Psi\,\e^{i\,\Theta(x)}, \quad 
A_{\mu} \to A_{\mu}-\frac{1}{e}\,\partial_{\mu}\,\Theta, \quad 
\bar{\Psi} \to \bar{\Psi} 
\e^{-i\,\Theta(x)}.
\label{eq22}
\end{equation}
Using the freedom of gauge invariance, we choose the axial gauge $e\,A_0=V(x)$, $A_{1}=0$. With $\bar{\Psi}=\Psi^{\dagger}\,\gamma^{0}$, we then have 
\begin{equation}
\mathcal{L}_{3} = -V(x)\,\Psi^{\dagger}\,\Psi=-V(x)\,[|u(x,t)|^2+|v(x,t)|^2].
\label{eq24}
\end{equation}
The corresponding nonlinear equations for the spinor components are:
\begin{eqnarray}
	\label{eq69a}
	i\,(u_t-u_x)+v+u^\star\,v^2-V(x)\,u&=&0,
	\\
	i\,(v_t+v_x)+u+v^\star\,u^2-V(x)\,v&=&0.
	\label{eq69b}
\end{eqnarray}

Conservation laws for this spinor model with an external field can be derived using the same method as described in Sec. \ref{sec2}. The resulting continuity equation for the charge density is exactly the same as Eq.\,(\ref{eq5a}). Therefore, in this system the charge 
\begin{equation}
	\label{eqQnew}
	Q=\int_{-\infty}^{+\infty} dx\,\rho(x,t)=\int_{-\infty}^{+\infty} dx\, \left(|u(x,t)|^2+|v(x,t)|^2\right)
\end{equation}
 is also conserved. 

However, there appears a source term in the continuity equation for the momentum density 
\begin{equation}
     \mathcal{P}_t+\Phi_x=-V(x) \frac{d}{dx} \rho(x). 
\end{equation}
Thus, in general, the momentum 
\begin{equation}
	\label{eqPnew}
	 P=\int_{-\infty}^{+\infty} dx\,\mathcal{P}(x,t)=\int_{-\infty}^{+\infty} dx\, \frac{i}{2}\,[u_x^\star\,u-u_x\,u^\star+v_x^\star\,v-v_x\,v^\star]
\end{equation}
will not be a conserved  quantity. 

Finally, the energy density is now
\begin{eqnarray}
	\label{eq60}
	T^{00}&=&\frac{i}{2}\,\left[\bar{\Psi}\,\gamma^{0}\,\partial^{0}\,\Psi-\partial^{0}\,
	\bar{\Psi}\,\gamma^{0}\,\Psi\right]-\tilde{\mathcal{L}}.
\end{eqnarray}
Here the first two terms cancel with the terms for $\mu=0$ in $\tilde{\mathcal{L}}$. Consequently, 
\begin{eqnarray}
	\label{eq61}
	T^{00}&=&-\frac{i}{2}\left[\bar{\Psi}\,\gamma^{1}\,\partial_{x}\,\Psi-\partial_{x}\,
	\bar{\Psi}\gamma^{1}\Psi\right]-\bar{\Psi}\,\Psi-\frac{1}{2}(\bar{\Psi}\,\Psi)^2+\frac{1}{4}\,J_{\mu} J^{\mu}
	-\mathcal{L}_3. \nonumber \\
	&=&\frac{i}{2}\left[u_xu^{\star}-u_x^{\star}u-v_xv^{\star}+v_x^{\star}v \right]
	-(uv^{\star}+u^{\star}v) \nonumber
	\\ 
	&&-\frac{1}{2}\left[(uv^{\star})^2+(u^{\star}v)^2 \right]
	+V(x)\left[|u|^2+|v|^2\right]. \label{eq62}
\end{eqnarray}
Note that $T^{00}$ is the energy density $\mathcal{H}$ in Eq.\ (\ref{eq7b}) plus  the additional term  $V(x) \rho(x)$.  Moreover, $T^{00}$ fulfills the continuity equation
 \begin{equation}
 	\label{eqT00}
 	T^{00}_t+\mathcal{J}_x=0,
 \end{equation}
where the energy current density $\mathcal{J}$ is the same as in Eq.\,(\ref{eq7c}). 
Therefore, the soliton energy in the external field 
 \begin{equation}
 	\label{eqEnew}
\tilde E=\int_{-\infty}^{+\infty}dx\,T^{00}(x,t)
\end{equation}
is conserved.

\section{ Variational ansatz and collective coordinates theory} \label{sec4}

An approximate solution for a moving soliton subjected to the potential $V(x)$ can be constructed using a  trial wave function with the same functional form as the exact solution Eqs. 
(\ref{eq20a})-(\ref{eq20b})
\begin{eqnarray} \label{eq20aa}
u(x,t) &=& \e^{-\beta/2} \,a[\gamma (x-v_s t)]\,\e^{i\,\theta[\gamma (x-v_s t)]}\,\e^{-i\,\omega\,\gamma(t-v_s x)},\\ 
\label{eq20bb}
v(x,t) &=& -\e^{\beta/2} \,a[\gamma (x-v_s t)]\,\e^{-i\,\theta[\gamma (x-v_s t)]}\,\e^{-i\,\omega\,\gamma(t-v_s x)},
\end{eqnarray}
with $\gamma=\cosh(\beta)$, where $\beta$ is the rapidity. 
 The trial function depends on time through the so-called collective coordinates (CCs) describing the position $q(t)$, rapidity $\beta(t)$,  and momentum $p(t)$ of the soliton, that is we replace 
\begin{eqnarray}
v_s\,t &\to& q(t), \quad \beta \to \beta(t),  \quad  
\gamma\,\omega\,v_s \to p(t), \quad \omega\,\gamma\,t \to p(t)\,q(t).
\label{eq25}
\end{eqnarray}
 The term $p(t) q(t)$  is needed in order to make the substitution 
%\begin{equation}
$\gamma (x-q)=z$ 
%\end{equation}
in the  trial wave function  
\begin{eqnarray}
u(x,t) &=& \e^{-\beta/2}\,a(z)\,\e^{i\,\theta(z)}\,\e^{i\,p\,(x-q)} \label{eq26a} \\
v(x,t) &=& -\e^{\beta/2}\,a(z)\,\e^{-i\,\theta(z)}\,\e^{i\,p\,(x-q)}
\label{eq26b}
\end{eqnarray} 
with $z=\cosh(\beta)\,(x-q)$. The variable $\omega$ is not treated as a collective coordinate because, as we have shown in  Sec.~\ref{sec3}, here the charge $Q(\omega)$ is conserved. 
This ansatz will be justified a posteriori by comparing the results of the CC theory with simulations (Section \ref{sec5}).

 When we now insert our ansatz into the Lagrangian 
(\ref{eq23a}), the integration over $x$ is replaced by an integration over $z$ 
and we obtain for $L=L_0+L_1+L_2+L_3$
\begin{eqnarray}
L_0&=&\int_{-\infty}^{+\infty} dx\, \frac{i}{2}\,[(u_t-u_x)\,u^\star-(u_t^\star-u_x^\star)\,u+(v_t+v_x)\,v^\star
-(v_t^\star+v_x^\star)\,v]  \nonumber \\
&=& I_0\,\dot{q}\,\sinh(\beta)+Q\,p\,\dot{q}-I_0\,\cosh(\beta)-Q\,p\,\tanh(\beta),
\end{eqnarray}
\begin{equation}\label{eq32}
	L_1=\int_{-\infty}^{+\infty} dx\,(u\,v^\star+u^\star\,v)=-\frac{I_1}{\cosh(\beta)},
\end{equation}
\begin{equation}\label{eq34}
	L_2=\frac{1}{2}\,\int_{-\infty}^{+\infty} dx\,
	[(u\,v^\star)^2+(u^\star\,v)^2]=\frac{I_2}{\cosh(\beta)},
\end{equation}
and
\begin{eqnarray} \nonumber
	L_3&=&-\,\int_{-\infty}^{+\infty} dx\,
	V(x)\,[|u|^2+|v|^2]=-\,\int_{-\infty}^{+\infty}\,dz\,V\left(\frac{1}{\gamma}z+q\right)\,
	\frac{\rho(z)}{\gamma} \\
	&=&-U(q,\beta), \label{eq36}
\end{eqnarray}
where $Q$ is given by Eq.\ (\ref{eq12}), and we have defined the integrals
\begin{eqnarray} \label{eq31}
I_0(\omega)&=&-\int_{-\infty}^{+\infty} dz\,\rho(z)\,\theta'(z)\, , \\
I_1(\omega)&=&\int_{-\infty}^{+\infty} dz\,\rho(z)\,\cos[2\,\theta(z)], \label{eq33} \\
\label{eq35}
I_2(\omega)&=&\frac{1}{4}\,\int_{-\infty}^{+\infty} dz\,\rho^2(z)\,\cos[4\,\theta(z)], 
\end{eqnarray}
which fulfill the relationships 
\begin{equation}
I_2=I_0,	\quad I_1=I_0+\omega Q,
	\label{eqI}
\end{equation} 
and can be calculated analytically  from Eq.\ (\ref{eq12}) and from  
\begin{eqnarray} \label{eqI0}
	I_0(\omega)&=&\sqrt{2}\,\ln\frac{1+\sqrt{2}\,\kappa}{1-\sqrt{2}\,\kappa}-2\, \omega  \, \ln\left[\frac{\omega+\kappa}{\omega-\kappa}\right].
\end{eqnarray}

Thus, the Lagrangian in terms of the CCs is 
\begin{eqnarray}
L&=&Q\,[p\,\dot{q}-p\,\tanh(\beta)]-I_0\,[\cosh(\beta)-\dot{q}\,\sinh(\beta)]-\frac{I_1}{\cosh(\beta)}+ \nonumber \\
&+&
\frac{I_2}{\cosh(\beta)}-U(q,\beta).\label{eq38}
\end{eqnarray}
 The above expression is the same as the Lagrangian of the Gross-Neveu equation with an external field \cite{mertens:2012}, except for an irrelevant phase $\phi(t)$:
\begin{eqnarray}
	L_{GN}&=&\tilde{Q}\,[p\,\dot{q}+  \dot{\phi}-p\,\tanh(\beta)]-\tilde{I}_0\,[\cosh(\beta)-\dot{q}\,\sinh(\beta)]-\frac{\tilde{I}_1 }{\cosh(\beta)}+ \nonumber \\
	&+&
	\frac{\tilde{I}_2}{2 \cosh(\beta)}-\tilde{U}(q,\beta).\label{eqLGN}
\end{eqnarray} 
The term $\dot{\phi}$ comes from a non-optimal choice of the CC in Ref. \cite{mertens:2012}, and has no influence on the CC equations of motion.  The expressions for the integrals $Q$, $I_0$, $I_1$, $I_2$, and $U$ in Eq.\ (\ref{eq38})  naturally differ from those given  in Ref. \cite{mertens:2012}  for $\tilde{Q}$, $\tilde{I}_0$, $\tilde{I}_1$, $\tilde{I}_2$, and $\tilde{U}$.
Indeed, $\tilde{Q}$, $\tilde{I}_{0}$, $\tilde{I}_{1}$
and $\tilde{U}$ in the Gross-Neveu and $Q$, $I_{0}$, $I_{1}$ and $U$
in the ABS models have the same functional dependence on the spinor
components, respectively, although the specific expressions for the
spinor field $\Psi$ differ in both systems. On the contrary, owing
to the nonlinear term, the integrals $I_2$ and $\tilde{I}_{2}$ are
different functionals of the spinor components. However, since the
identities $I_2=I_1+\omega Q$ and $\tilde{I}_2=\tilde{I}_1+\omega
\tilde{Q}$ are satisfied, the differences in the nonlinear term do 
not lead to  drastic changes in the Lagrangian $L$.  This result is of major interest, because it means that the soliton dynamics in the ABS spinor model is \textit{qualitatively the same} as in the  Gross-Neveu equation, although the nonlinearities of the two systems are very different. This is in contrast to the nonlinear Schr\"odinger (NLS) equation where the soliton dynamics is \textit{qualitatively different} for different types of the nonlinearity \cite{cooper:2012,dawson:2017}.

 The Lagrange equation for $q(t)$ yields
\begin{equation}
	\dot{P}_q=-\frac{\partial U}{\partial q}, 
	\label{eq41}
\end{equation} 
where the canonical momentum 
\begin{equation}
P_q=\frac{\partial L}{\partial \dot{q}}= Q\,p+I_0\,\sinh(\beta) 
\label{eq40}
\end{equation} 
is conjugated to the soliton position $q$. Thus, the time derivative of the momentum is equal to an external force. 

The Lagrange equation for $p(t)$ results in 
\begin{equation}
\dot{q}=\tanh(\beta), \quad  
\label{eq42}
\end{equation}
with $\gamma=\frac{1}{\sqrt{1-\dot{q}^2}}=\cosh(\beta)$, 
which agrees with Eq.\ (\ref{eq16}).

From the Lagrange equation for $\beta(t)$, we obtain, after several calculations 
\begin{equation}
Q\,p= \gamma\,\dot{q}\,(I_1-I_2)-\frac{\partial U}{\partial \dot{q}}.  
\label{eq43}
\end{equation}
Here the identity
\begin{equation}
\frac{\partial U}{\partial\beta}=\frac{\partial U}{\partial \dot{q}}\, \sech^2(\beta) \label{eq44}
\end{equation} 
is used. 

 Note that Eq.\ (\ref{eq43}) can be combined with Eq.\ (\ref{eq40}) in order to eliminate $p(t)$. In this way, we finally obtain a Newtonian-like equation of motion
\begin{equation}
\frac{d}{dt}\,(M\,\dot{q})=F_{eff}(q,\dot{q}), 
\label{eq45}
\end{equation} 
where $M=\gamma\,M_0$ is a relativistic mass, and 
\begin{equation}
M_0=I_0+I_1-I_2=I_1=\sqrt{2}\,\ln\left[\frac{1+\sqrt{2}\,\kappa}{1-\sqrt{2}\,\kappa}\right] 
\label{eq46}
\end{equation}
is the rest mass, and the effective force reads
\begin{equation}
F_{eff}(q,\dot{q})=\frac{d}{dt}\,\frac{\partial U}{\partial \dot{q}}-\frac{\partial U}{\partial q}.  
\label{eq47}
\end{equation} 
This means that the soliton behaves like a relativistic particle, driven by an effective force. 
Therefore, the CC approximation reduces a problem with an infinite number of 
degrees of freedom to only one second-order ordinary differential equation. Once Eq.\ (\ref{eq45}) is solved the rapidity $\beta(t)$ and the momentum $p(t)$  
can be immediately  obtained by Eqs.\  (\ref{eq42}) and  (\ref{eq43}). 
%Finally, $\phi(t)=\omega_0\,\gamma(\dot{q}(t))\,t-p(t)\,q(t)$, where $\omega_0=\omega(0)$ is fixed %by initial frequency at $t=0$. 

Futhermore, inserting the ansatz in Eqs.~(\ref{eqQnew}), (\ref{eqPnew}),  and (\ref{eqEnew}) we can obtain explicit approximate expressions for the charge, the momentum and the soliton energy, respectively. For the charge it turns out 
\begin{equation}
	Q = 2\, \ln\left[\frac{\omega+\kappa}{\omega-\kappa}\right]. 
	\label{eq55} 
\end{equation}
And for the momentum
\begin{equation}
	P=I_0\,\sinh(\beta)+Q\,p \, ,
	\label{eq49}
\end{equation} 
which agrees with the canonical momentum $P_q$ in Eq.\ (\ref{eq40}). Taking into account Eqs.~(\ref{eq43}) and (\ref{eqI}) we obtain
\begin{equation}
	\label{eq68}
	 P=\gamma\,M_0\,\dot{q}-\frac{\partial U}{\partial \dot{q}}.
\end{equation}
Finally for the energy we have
\begin{equation}
	\label{eq63}
	\tilde E=I_0\,\cosh(\beta)+Q\,p\,\tanh(\beta)+\frac{I_1}{\cosh(\beta)}-\frac{I_2}{\cosh(\beta)}+U(q,\dot{q}), 
\end{equation}
which, using Eqs.\ (\ref{eq43}),  (\ref{eqI}), and (\ref{eqI0}) can be expressed as
\begin{equation}
	\label{eq64}
	\tilde E=\gamma\,M_0+U(q,\dot{q})-\dot{q}\,\frac{\partial U}{\partial \dot{q}}.
\end{equation}
In particular, the rest energy is 
\begin{equation}
	\label{eq65}
	\tilde E_0=M_0+U(q,0).
\end{equation}  
It is important to note that, using the CC equations of motion, one can prove that $\frac{d\tilde E}{dt}=0$, which is consistent with the conservation of the soliton energy.

\section{Collective coordinates theory compared to simulations} \label{sec5}

In this section the equation of motion (\ref{eq45}) of the CC-theory is solved analytically or numerically and compared with simulations, i.e. numerical solutions of the  ABS nonlinear spinor model with a potential $V(x)$, equations  (\ref{eq69a})-(\ref{eq69b}). In order to solve them, we use a Runge-Kutta-Verner fifth-order algorithm with variable time step and a spectral method for the computation of spatial derivatives \cite{cuevas:2015}. The system is discretized by taking constant spatial intervals $\Delta x=0.02$, and a sufficiently large number of points $N$  such that the length of the system is much longer than the soliton width. 

As initial condition, we take the trial wave function Eqs. (\ref{eq26a})-(\ref{eq26b}) at $t=0$ %$x'=\gamma_0\,x$ and $t'=-\gamma_0\,v_0\,x$:
\begin{eqnarray}
\label{eq70a}
u(x,0)&=& e^{-\beta_0/2}\,a(\gamma_0\,[x-q_0])\,e^{i\,\theta(\gamma_0\,[x-q_0])}\,e^{i\,p_0\,(x-q_0)},
\\
v(x,0)&=& -e^{\beta_0/2}\,a(\gamma_0\,[x-q_0])\,e^{-i\,\theta(\gamma_0\,[x-q_0])}\,e^{i\,p_0\,(x-q_0)},
\label{eq70b}
\end{eqnarray}
where $\gamma_0=1/\sqrt{1-v_0^2}$, $v_0=\dot{q}(0)$, $q_0=q(0)$, $p_0=\omega\,\gamma\,v_0$, and  $\beta_0=$arctanh$(v_0)$. During the simulations, after computing the spinor components, we calculate the soliton position by means of 
\begin{equation}
\label{eq71}
q(t)=\frac{Q_1(t)}{Q}=\frac{{\int\,dx\,x\,\rho(x,t)}}{\int\,dx\,\rho(x,t)},
\end{equation}
and also the  momentum and the soliton energy using Eqs.~(\ref{eqPnew}) and (\ref{eqEnew}) respectively. 

 In our study three simple potentials are going to be considered: 

 \subsection{Ramp with negative slope} \label{rampneg}
 
 \begin{equation}
 \label{eq72}
 V(x)=-V_1\,x, \qquad V_1>0.
 \end{equation}
 The effective potential Eq.\ (\ref{eq36}) is 
\begin{equation}
\label{eq73}
U(q)=\int_{-\infty}^{+\infty}dz\,V\left(\frac{z}{\gamma}+q\right)\,\frac{\rho(z,t)}{\gamma}=-V_1\,Q\,q(t)  \, .
\end{equation} 
So the equation of motion Eq.\ (\ref{eq45}) reads
\begin{equation}
\label{eq74}
\frac{d}{dt}(\gamma\,M_0\,\dot{q})=V_1\,Q \, , 
\end{equation} 
whose general solution is
\begin{equation}
\label{eq75}
q(t)=\frac{1}{C_1}\,\sqrt{(C_1\,t+C_2)^2+1}-\frac{1}{C_1}\,\sqrt{C_2^2+1}+q(0) \, ,
\end{equation}
where
\begin{equation}
\label{eq76}
C_1=\frac{V_1\,Q}{M_0}, \quad C_2=\frac{\dot{q}(0)}{\sqrt{1-[\dot{q}(0)]^2}}.
\end{equation}
For $v_0=\dot{q}(0)=0$ we obtain 
\begin{equation}
\label{eq77}
q(t)=\frac{1}{C_1}\,[\sqrt{C_1^2\,t^2+1}-1]+q(0) \, .
\end{equation}
 Moreover, from Eqs.\ (\ref{eq38}) and (\ref{eq74}), it also follows that the momentum is linear in $t$: 
\begin{equation}
\label{eqmome}
P=\gamma\,M_0\,\dot{q}= \ V_1\,Q\,t.
\end{equation}

\begin{figure}[ht!]
	\centering
	\begin{tabular}{cc}
		\includegraphics[width=0.4\linewidth]{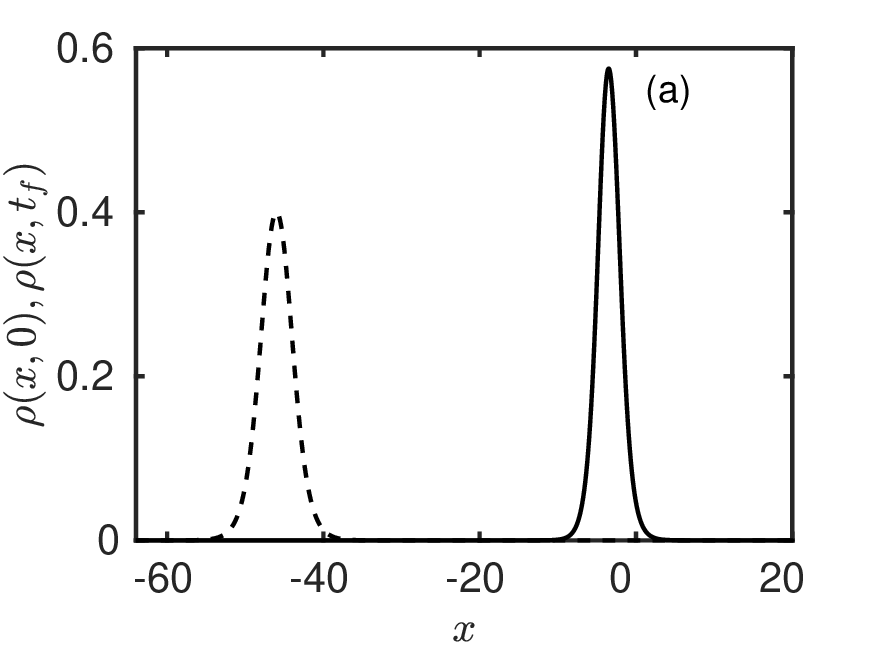} &
		\includegraphics[width=0.4\linewidth]{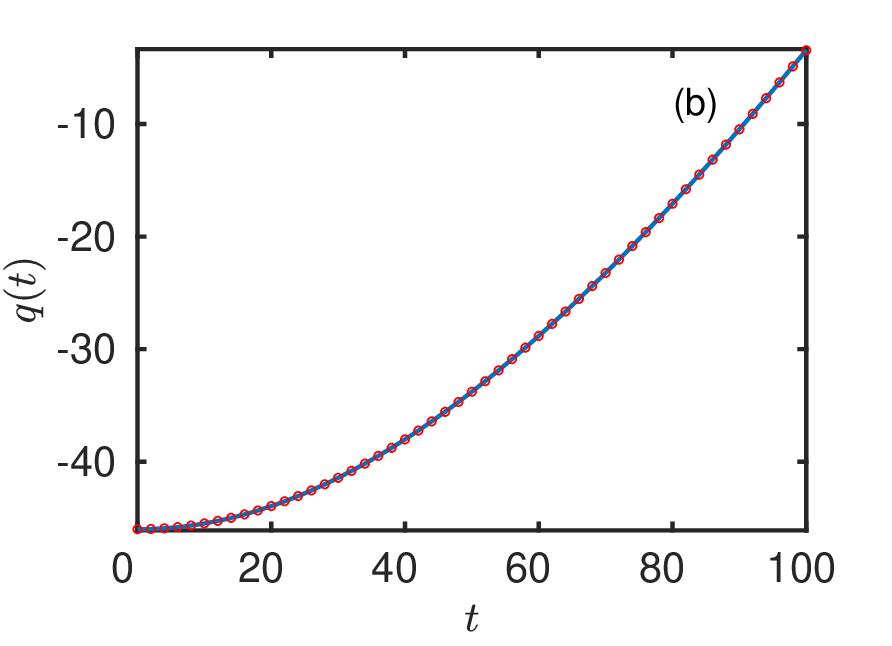} \\
		\includegraphics[width=0.4\linewidth]{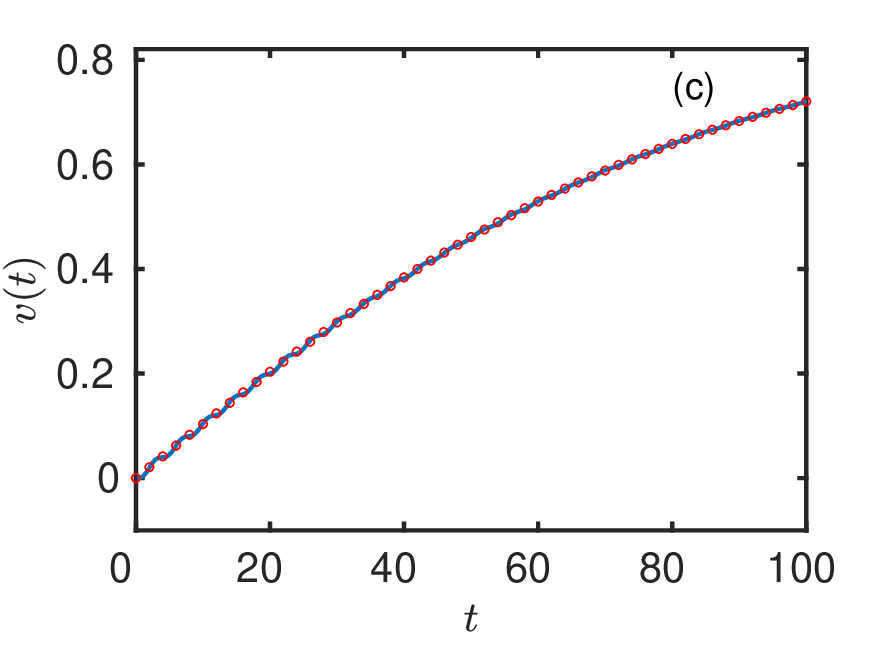} &
		\includegraphics[width=0.4\linewidth]{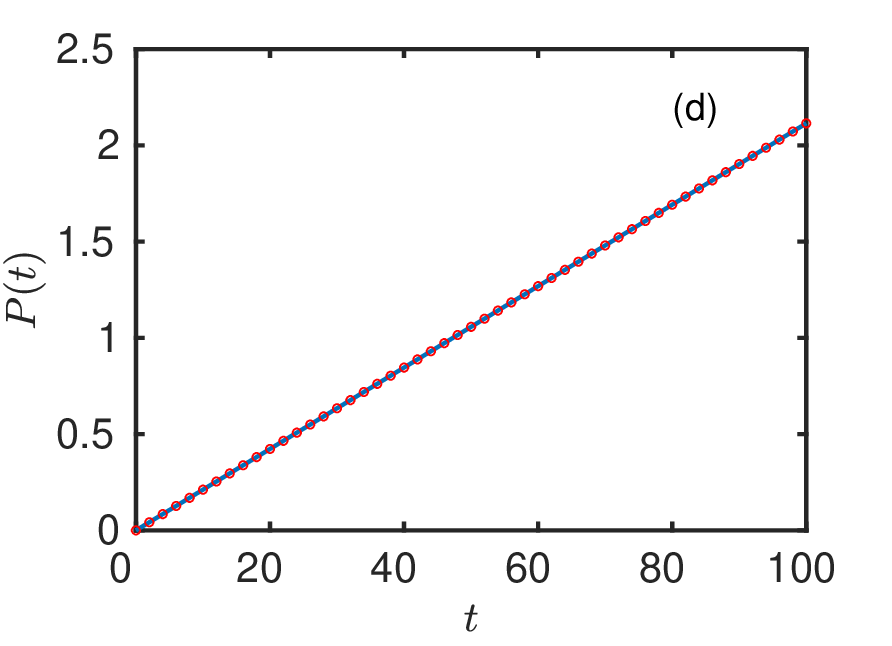} 
	\end{tabular}
	\caption{Soliton dynamics with the ramp potential $V(x)=-V_1\,x$, for $\omega=0.9$ and $V_1=0.01$. Panel (a): soliton profiles at $t=0$ (dashed line) and at $t_f=100$ (solid line). 
	Panels (b), (c) and (d): comparison of the simulation results for the soliton position $q(t)$, velocity $v(t)$, and momentum $P(t)$ (solid lines) with the CC results (red circles). 	
	 Initial conditions: $q(0)=-46$, and $v(0)=0$. } 	
	\label{fig1}
\end{figure}

 For Fig.\ \ref{fig1} we have chosen $\omega=0.9$, which lies in the interval $[0.75,1.0)$. In panel (a) it can be observed that the charge density is bell-shaped, Lorentz-contracted and its height increases, as predicted by Eq.~(\ref{eqrho}). The charge and the energy are constants and agree with Eq.~(\ref{eq55}) and Eq.~(\ref{eq64}), respectively. The soliton position $q(t)$, shown with circles in panel (b), 
velocity $v(t)$ (circles in panel (c)), and the momentum $P(t)$ (circles in panel (d)) agree perfectly with the CC results plotted with lines, that is, the CC curves in Fig.\ \ref{fig1} superimpose the simulation results.

\begin{figure}[ht!]
	\centering
	\begin{tabular}{cc}
		\includegraphics[width=0.4\linewidth]{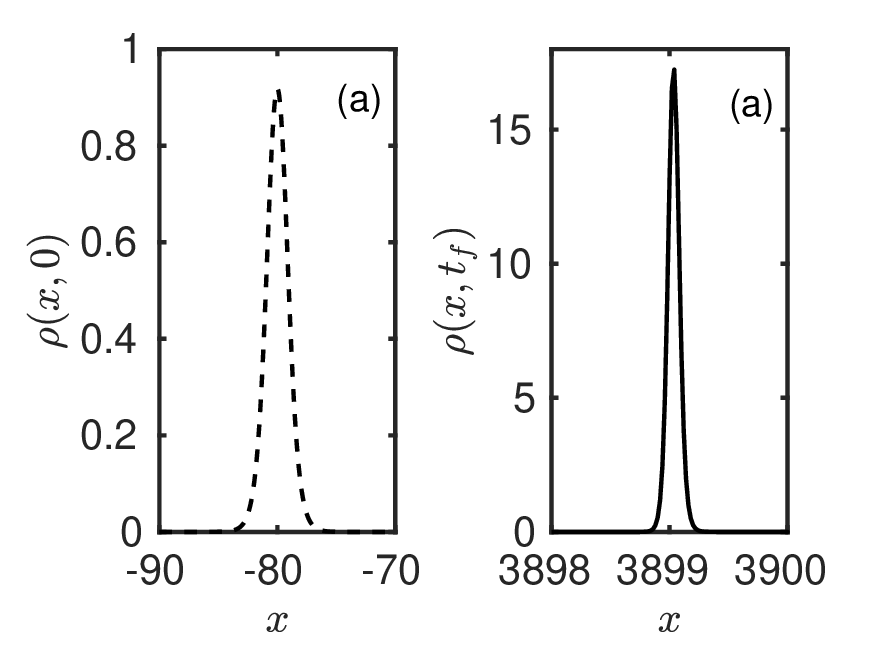} &
		\includegraphics[width=0.4\linewidth]{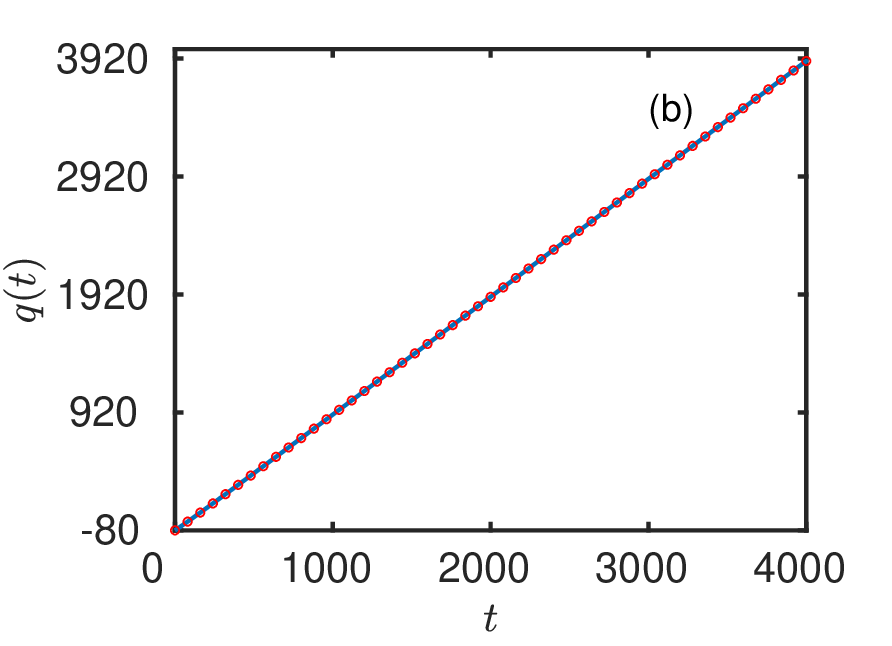} \\
		\includegraphics[width=0.4\linewidth]{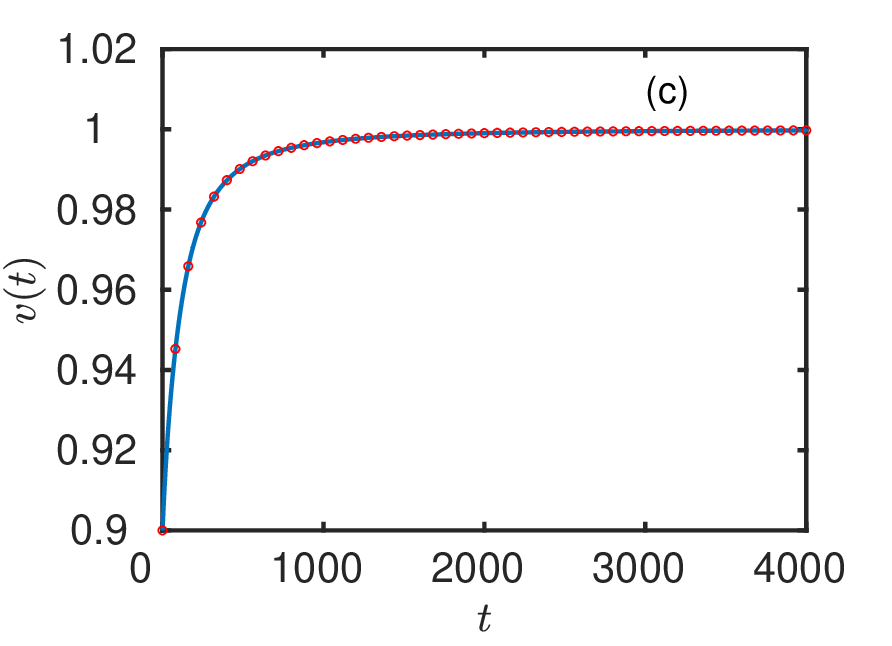} &
		\includegraphics[width=0.4\linewidth]{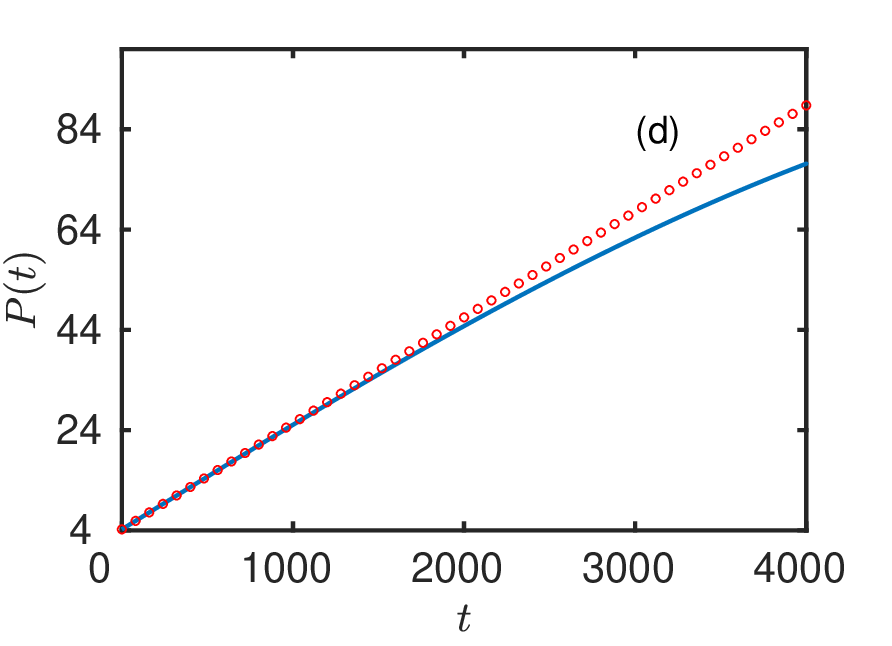} 
	\end{tabular}
	\caption{Soliton dynamics with the ramp potential $V(x)=-V_1\,x$ in the relativistic regime. Panel (a): soliton profiles at $t=0$ (dashed line) and at $t_f=4000$ (solid line).
	Notice that the width of the charge density  is Lorentz-contracted, while the height is increased such that the charge is conserved. 
	Panels (b), (c) and (d): comparison of the simulation results for the soliton position $q(t)$, velocity $v(t)$, and momentum $P(t)$ (solid lines) with the CC results (red circles).
	Initial conditions and parameters: $q(0)=-80$, $v(0)=0.9$, $\omega=0.9$, $V_1=0.01$.}
	\label{fig1A}
\end{figure}

In Fig.\ \ref{fig1A} we are in the relativistic  regime since we have taken $v(0)=0.9$. The position and the velocity (panels (b) and (c)) fully agree again with the results
of the CC theory, but the momentum $P(t)$ (see panel (d)) starts to differ for long times. This is due to the fact that, as a consequence of the Lorentz contraction, the soliton becomes very narrow and the computation of the spatial derivatives is less accurate.

\begin{figure}[ht!]
	\centering
	\begin{tabular}{cc}
		\includegraphics[width=0.4\linewidth]{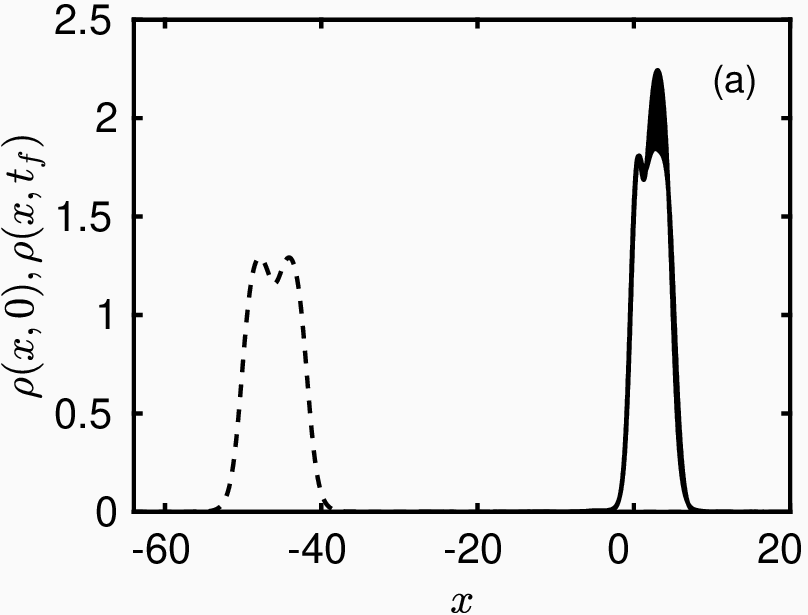} &
		\includegraphics[width=0.4\linewidth]{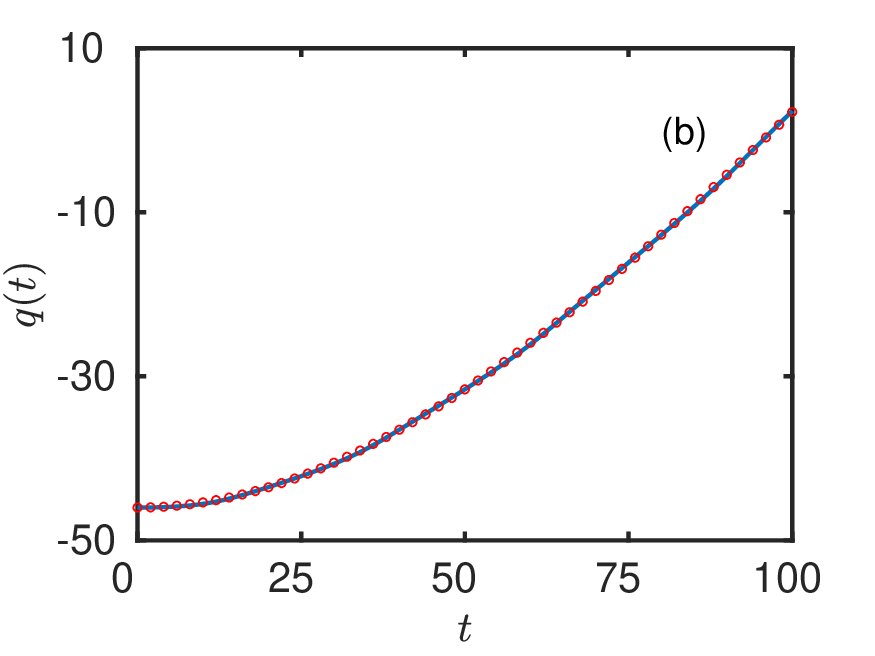} \\
		\includegraphics[width=0.4\linewidth]{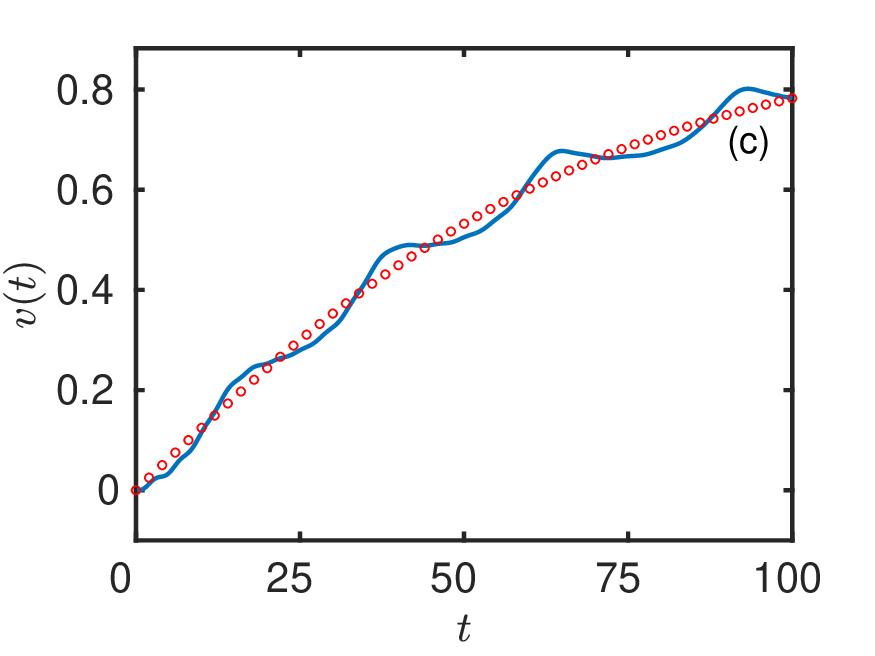} &
		\includegraphics[width=0.4\linewidth]{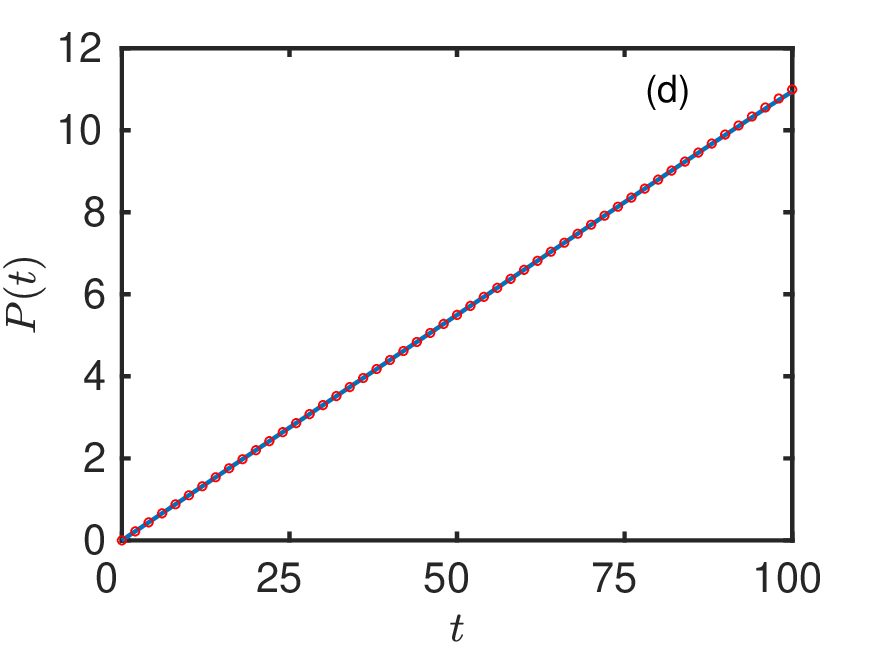} 
	\end{tabular}
	\caption{Soliton dynamics with the ramp potential $V(x)=-V_1\,x$, for $\omega=0.71$ and $V_1=0.01$. Panel (a): soliton profiles at $t=0$ (dashed line) and at $t_f=100$ (solid line). Note that the soliton is unstable. 
		Panels (b), (c) and (d): comparison of the simulation results for the soliton position $q(t)$, velocity $v(t)$, and momentum $P(t)$ (solid lines) with the CC results (red circles).
		Initial conditions: $q(0)=-46$, $v(0)=0$.}
	\label{fig2}
\end{figure}

For Fig.\ \ref{fig2}, we have chosen $v(0)=0$ and  $\omega=0.71$, which is situated near the lower end of the interval $1/\sqrt{2}<\omega<3/4$. Here the charge density has two humps and is slightly deformed during the time evolution, indicating that the soliton is unstable. 
Even so, $q(t)$ and $P(t)$ keep agreeing well with the CC results, but $v(t)$ shows small oscillations due to the increasing deformation of the soliton.

\subsection{Harmonic potential} \label{harmonic}

\begin{equation}
\label{eq76a}
V(x)=\frac{1}{2}\,V_2\,x^2=\frac{1}{2}\,\left(\frac{x}{l}\right)^2,
\end{equation}
where $l=1/\sqrt{V_2}$ is the characteristic length of the potential, which must be  
large compared to the soliton width $1/\kappa$. 

 For this choice
\begin{equation}
\label{eq77a}
U(q,\dot{q})=\frac{1}{2}\,V_2[Q\,q^2+I_3\,(1-\dot{q}^2)],
\end{equation}
with 
\begin{equation}
\label{eq78}
I_3=\int_{-\infty}^{+\infty}\,dz\,z^2\,\frac{\rho(z)}{\gamma}.
\end{equation}
The effective force Eq.\ (\ref{eq47}) is now
\begin{equation}
\label{eq79}
F_{eff}=-V_2\,I_3\,\ddot{q}-V_2\,Q\,q
\end{equation}
and the equation of motion
\begin{equation}
\label{eq80}
\frac{d}{dt}[\gamma\,M_0\,\dot{q}+V_2\,I_3\,\dot{q}] =-V_2\,Q\,q.
\end{equation}
In the non-relativistic regime $|\dot{q}^2| \ll 1$, $\gamma \approx 1$, we get the harmonic oscillator equation
\begin{equation}
\label{eq81}
\ddot{q}+\Omega^2\,q=0, \quad \Omega^2=\frac{V_2\,Q}{M_0+V_2\,I_3},
\end{equation}
i.e. the rest mass is increased by $V_2\,I_3$.
Choosing $q(0)=0$, $\dot{q}(0)=v_s$, we obtain
\begin{equation}
\label{eq82}
q(t)=\frac{v_s}{\Omega}\,\sin(\Omega\,t).
\end{equation} 

\begin{figure}[ht!]
	\centering
	\begin{tabular}{cc}
	\includegraphics[width=0.4\linewidth]{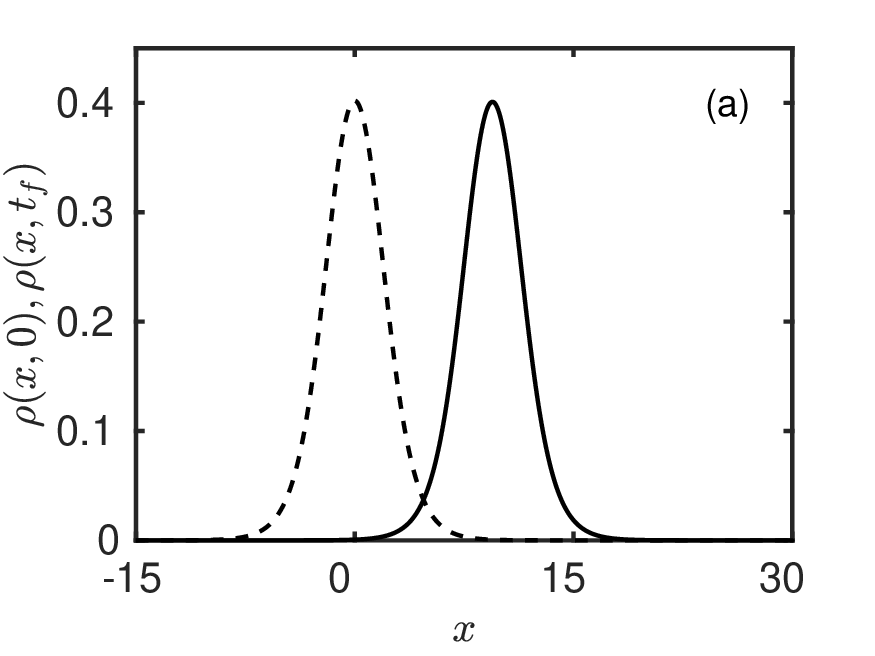} &
	\includegraphics[width=0.4\linewidth]{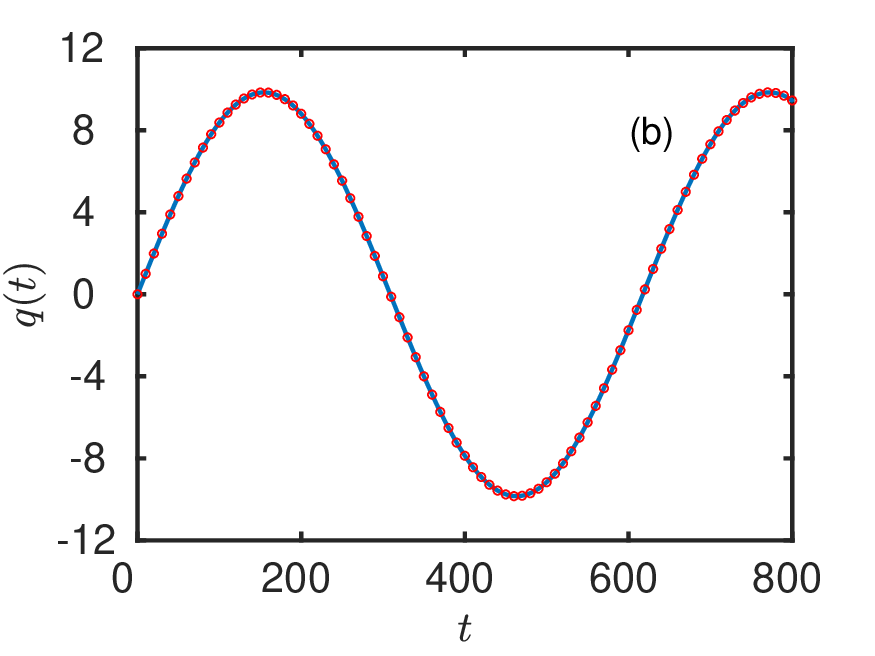} \\
	\includegraphics[width=0.4\linewidth]{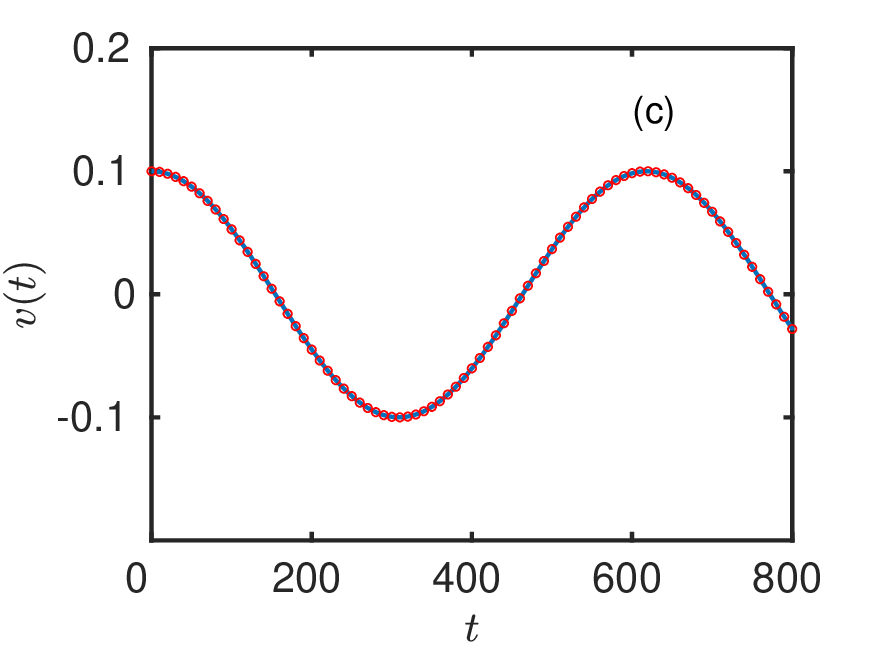} &
	\includegraphics[width=0.4\linewidth]{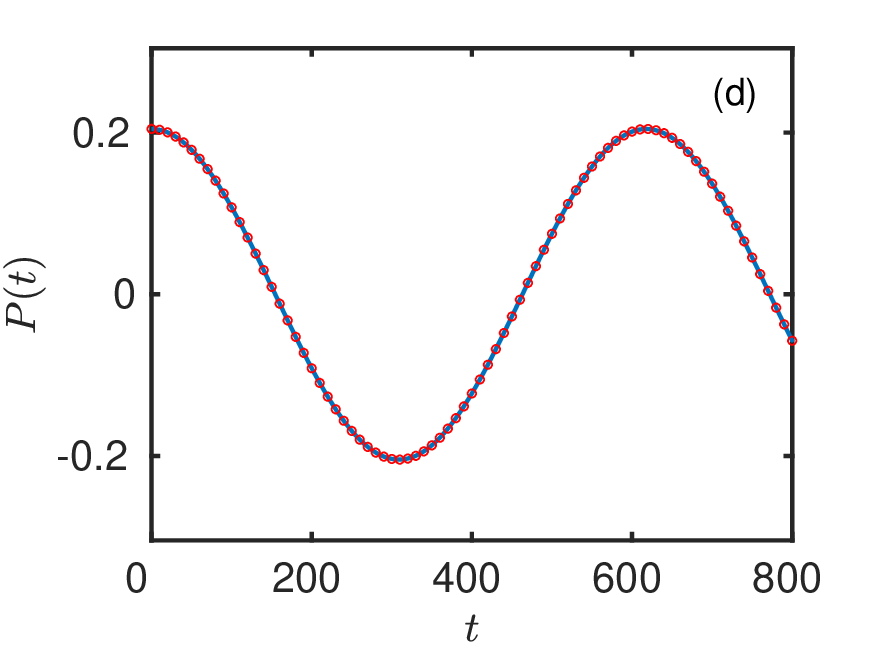} 
\end{tabular}
	\caption{Soliton dynamics with the harmonic potential $V(x)=\frac{V_2}{2}\,x^2$, for $\omega=0.9$ and  $V_2=0.0001$. Panel (a): soliton profiles at $t=0$ (dashed line) and at $t_f=800$ (solid line). 
		Panels (b), (c) and (d): comparison of the simulation results for the soliton position $q(t)$, velocity $v(t)$, and momentum $P(t)$ (solid lines) with the CC results (red circles).
		Initial conditions: $q(0)=0$, $v(0)=0.1$.}
	\label{fig3}
\end{figure}

In Fig.\ \ref{fig3}, for $\omega=0.9$, which lies in the upper interval $[0.75,1.0)$, we can see the  harmonic  oscillations performed by the soliton, in perfect agreement with the CC-results.
	
\begin{figure}[ht!]
	\centering
	\begin{tabular}{cc}
		\includegraphics[width=0.4\linewidth]{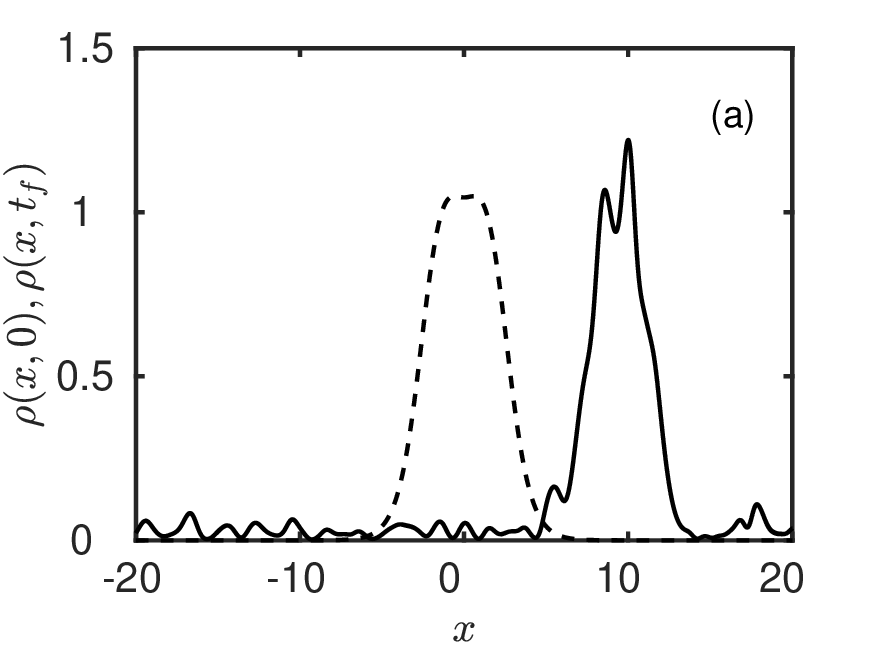} &
		\includegraphics[width=0.4\linewidth]{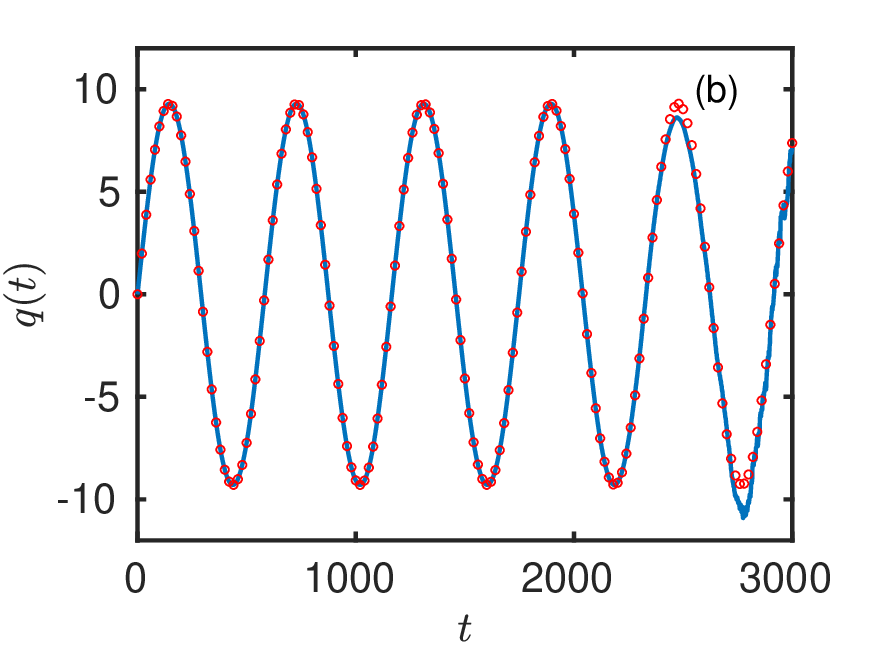} \\
		\includegraphics[width=0.4\linewidth]{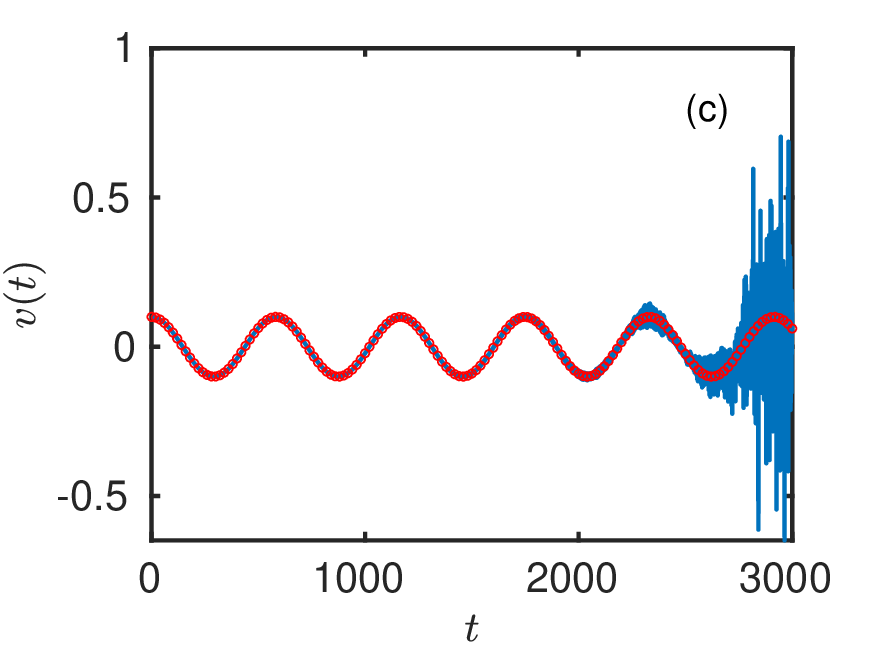} &
		\includegraphics[width=0.4\linewidth]{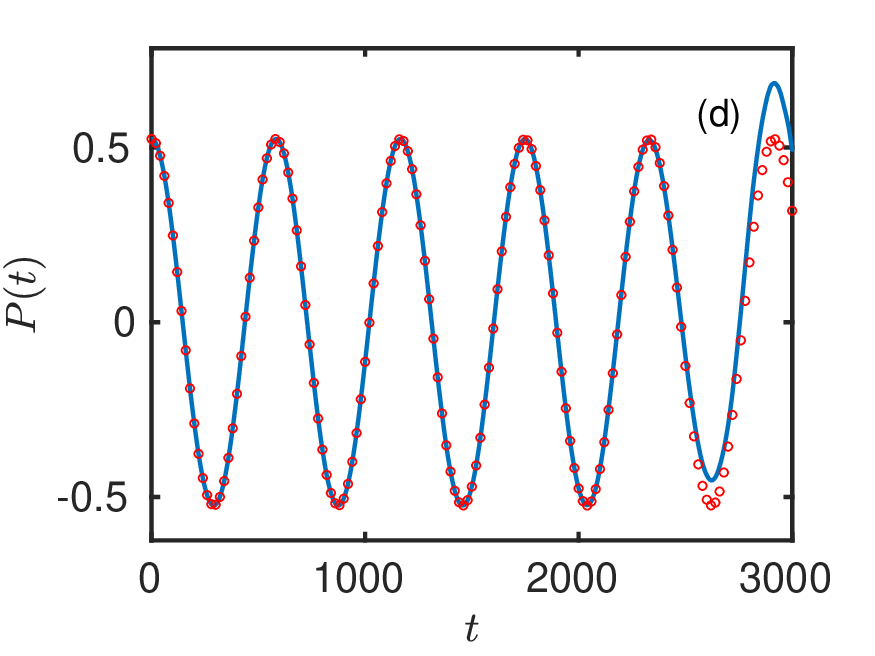} 
	\end{tabular}
	\caption{Soliton dynamics with the harmonic potential $V(x)=\frac{V_2}{2}\,x^2$, for $\omega=0.74$ and $V_2=0.0001$. Panel (a): soliton profiles at $t=0$ (dashed line)  and at $t_f=3000$ (solid line). The soliton is clearly unstable. 
		Panels (b), (c) and (d): comparison of the simulation results for the soliton position $q(t)$, velocity $v(t)$, and momentum $P(t)$ (solid lines) with the CC results (red circles). 
	Initial conditions: $q(0)=0$, $v(0)=0.1$.}
	\label{fig3a}
\end{figure}

If we choose $\omega=0.74$, which lies close to the \textit{upper end} of the interval $1/\sqrt{2}<\omega<3/4$, the soliton is stable for a rather long time, but eventually becomes unstable (see Fig.\ \ref{fig3a}). As in the ramp potential case,  when we take values of $\omega$ closer to the lower end of the above interval, the soliton is deformed faster and the instability appears earlier.

\begin{figure}[ht!]
	\centering
		\begin{tabular}{cc}
			\includegraphics[width=0.4\linewidth]{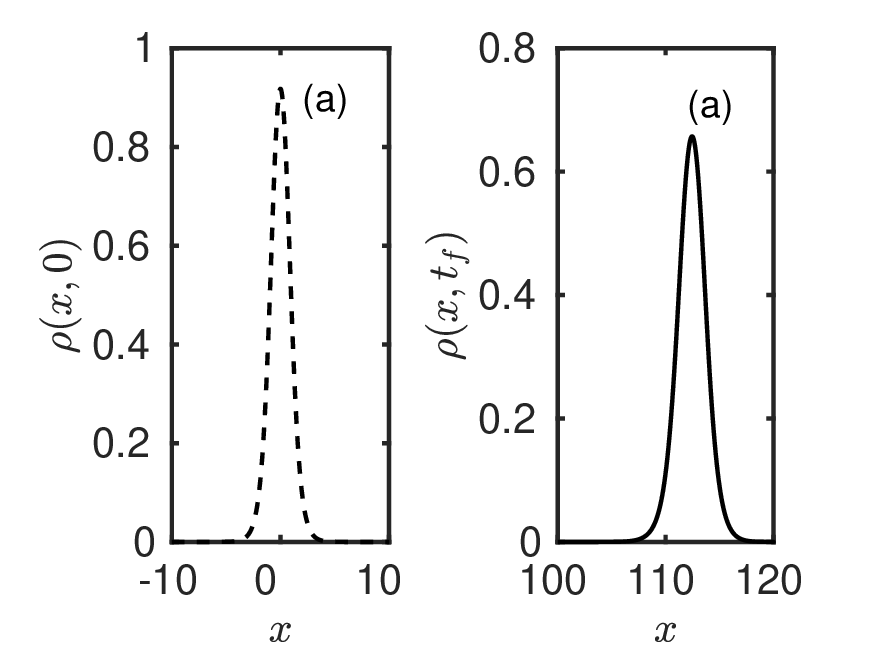} &
			\includegraphics[width=0.4\linewidth]{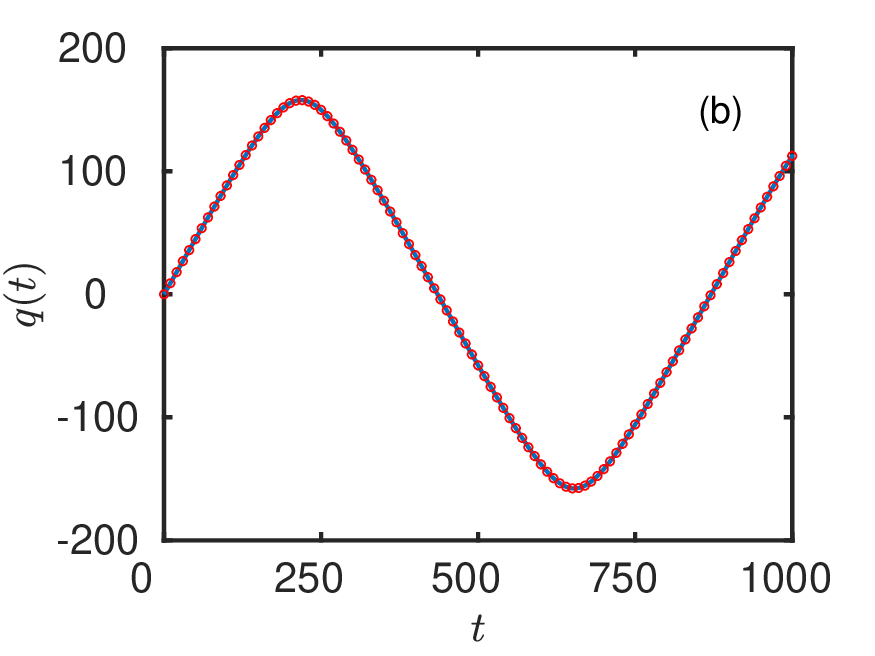}  \\
			 \includegraphics[width=0.4\linewidth]{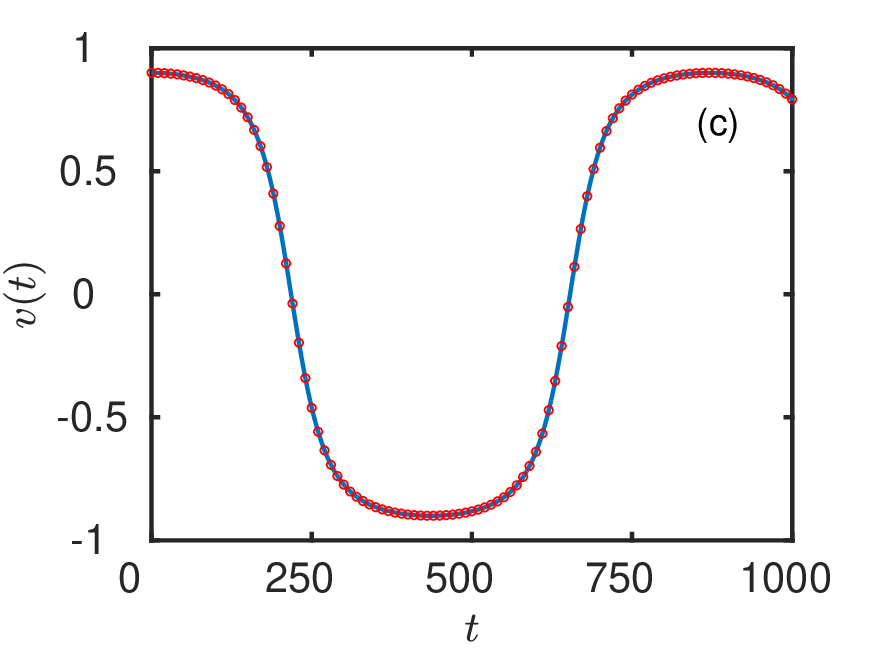} &
			 \includegraphics[width=0.4\linewidth]{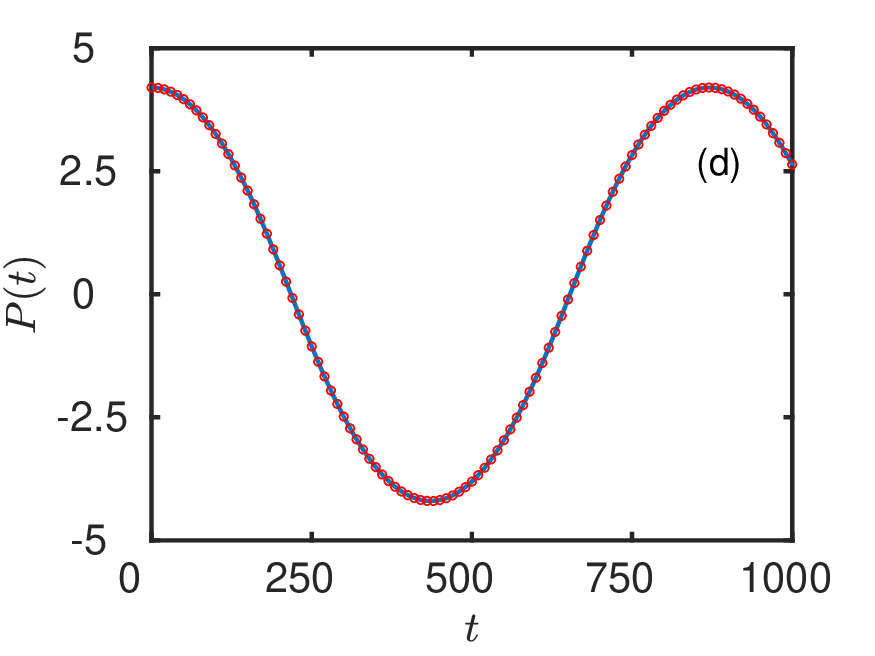}
		\end{tabular}
	\caption{Soliton dynamics with the harmonic potential $V(x)=\frac{V_2}{2}\,x^2$ in the relativistic regime. Panel (a): soliton profiles at $t=0$ (dashed line)  and at $t_f=1000$ (solid line). 
		Panels (b), (c) and (d): comparison of the simulation results for the soliton position $q(t)$, velocity $v(t)$, and momentum $P(t)$ (solid lines) with the CC results (red circles).  
	Initial conditions and parameters:   
		$q(0)=0$, $v(0)=0.9$, $\omega=0.9$, $V_2=0.0001$.}
	\label{fig5}
\end{figure}

In the relativistic regime, the equation of motion (\ref{eq80}) can be written as 
 \begin{equation}
 \label{eq83}
 (M_0\,\gamma^3+V_2\,I_3)\,\ddot{q} +V_2\,Q\,q=0
 \end{equation}
where $M_0\,\gamma^3$ is the so-called longitudinal relativistic mass. For $v(0)=0.9$, and $\omega=0.9$, $q(t)$ and $v(t)$ are strongly anharmonic as is shown in Fig.\ \ref{fig5}.  Nevertheless the soliton is stable and its dynamics agrees perfectly with the CC-result.
    
\subsection{Periodic potential} \label{periodic}

\begin{equation}
\label{eq84}
V(x)=-\epsilon\,\cos(k\,x), 
\end{equation}
where $\epsilon>0$. 
In this case, periodic boundary conditions in the simulations imply that the system length $L$ has to be an integer multiple of the spatial period $\lambda=2\,\pi/k$. Additionally $\lambda$ has to be commensurable with $\Delta x=0.02$.  We have taken $L=128=5^2 \cdot 2^8 \cdot \Delta x$. As a consequence, the values of the wave number $k$ used in the simulations are restricted to the discrete set $k_n=n\,\pi/64$, where $n$ is $5$, $25$ or a multiple of $2$.
 
\begin{figure}[ht!]
	\centering
	\begin{tabular}{cc}
	\includegraphics[width=0.4\linewidth]{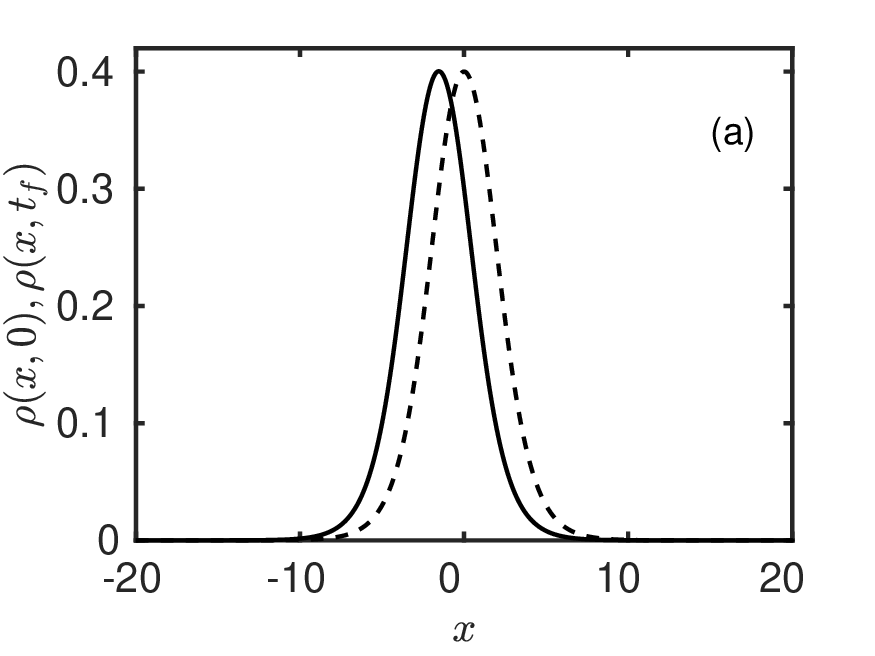} &
	\includegraphics[width=0.4\linewidth]{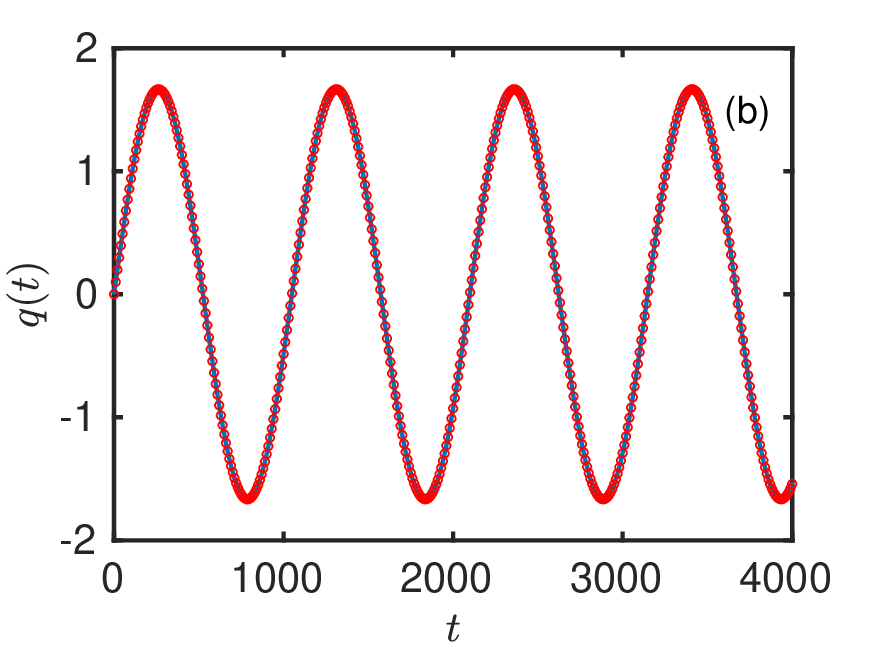} \\
	\includegraphics[width=0.4\linewidth]{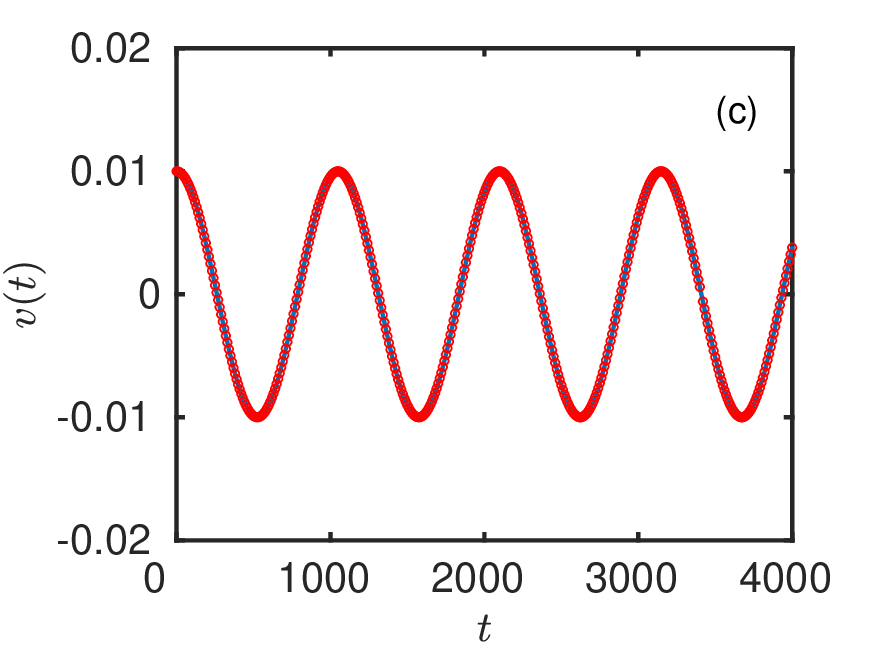} &
	\includegraphics[width=0.4\linewidth]{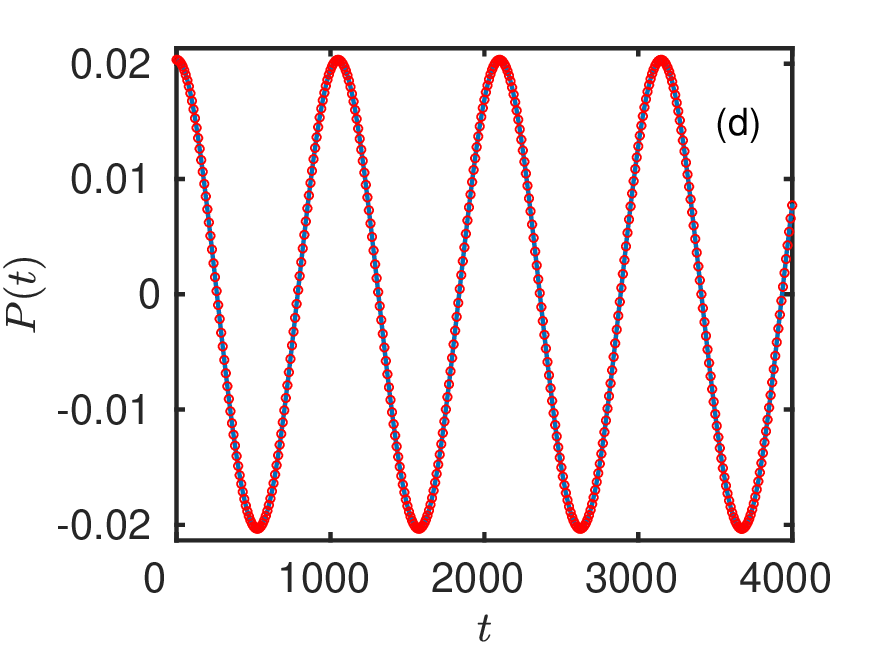} 
\end{tabular}	
	\caption{Soliton dynamics with the periodic potential $V(x)=-\epsilon\,\cos(k\,x)$, for $\omega=0.9$, $\epsilon=0.001$, and $k=\pi/16$.  Panel (a): soliton profiles at $t=0$ (dashed line)  and at $t_f=4000$ (solid line).  
		Panels (b), (c) and (d): comparison of the simulation results for the soliton position $q(t)$, velocity $v(t)$, and momentum $P(t)$ (solid lines) with the CC results (red circles).
		%Pannels (b), (c) and (d): Comparison of the soliton position $q(t)$, velocity $v(t)$, and momentum $P(t)$ (red circles) with the CC approximation (solid lines). 
		Results from the CC theory superimposed simulation results. 
		 Initial conditions:    
		$q(0)=0$, $v(0)=0.01 \ll v_c\simeq 0.06369$.}
	\label{fig6}
\end{figure}

\begin{figure}[ht!]
	\centering
		\begin{tabular}{cc}
		\includegraphics[width=0.4\linewidth]{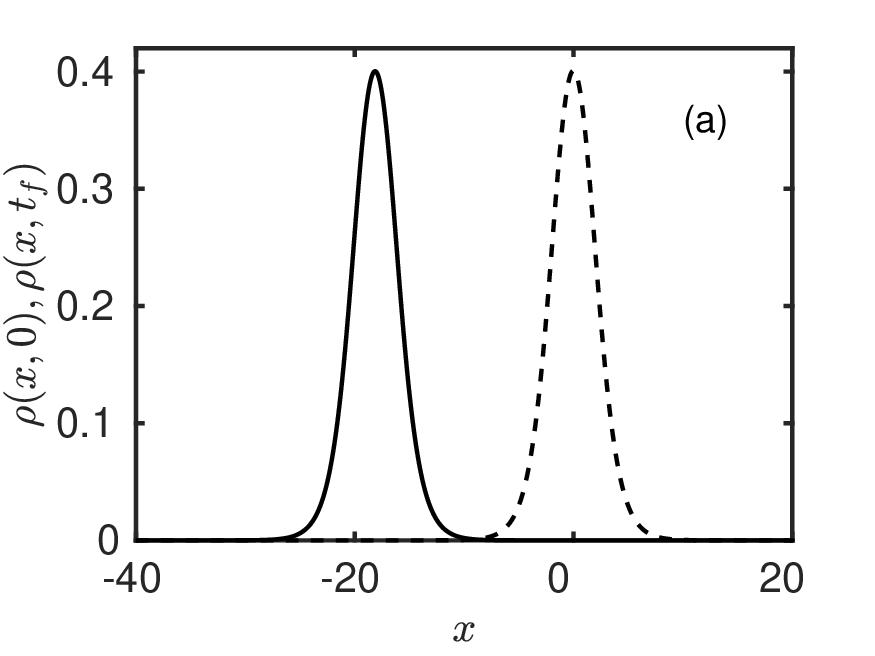} &
		\includegraphics[width=0.4\linewidth]{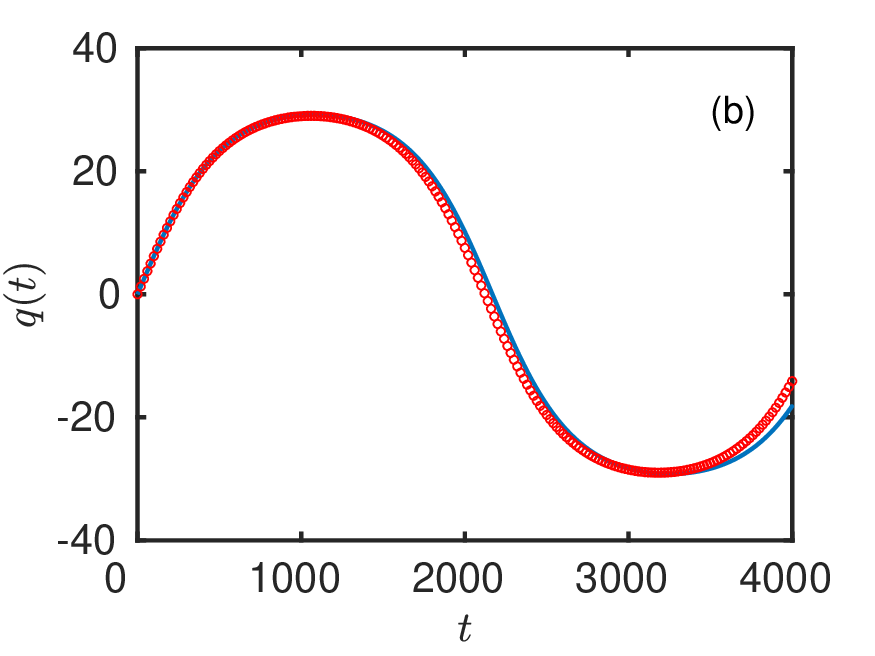} \\
		\includegraphics[width=0.4\linewidth]{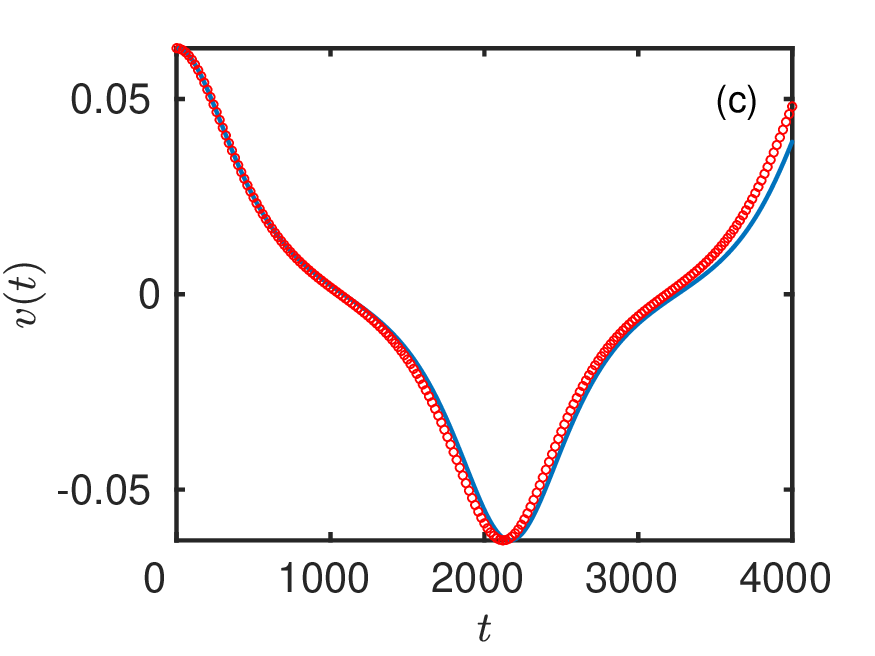} &
		\includegraphics[width=0.4\linewidth]{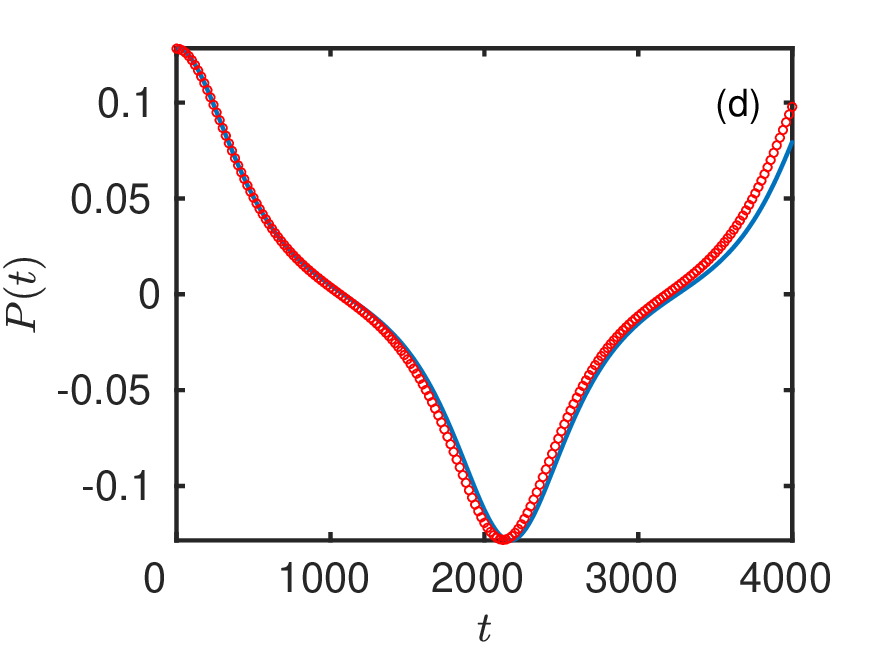} 
	\end{tabular}		
	\caption{Soliton dynamics with the periodic potential $V(x)=-\epsilon\,\cos(k\,x)$, for $\omega=0.9$, $\epsilon=0.001$, and $k=\pi/32$.  Panel (a): soliton profiles at $t=0$ (dashed line)  and at $t_f=4000$ (solid line).  
		Panels (b), (c) and (d): comparison of the simulation results for the soliton position $q(t)$, velocity $v(t)$, and momentum $P(t)$ (solid lines) with the CC results (red circles).
		Initial conditions:    
		$q(0)=0$, $v(0)=0.063<v_c\simeq 0.06369$.}
	\label{fig7}
\end{figure}

\begin{figure}[ht!]
	\centering
	\begin{tabular}{cc}
		\includegraphics[width=0.4\linewidth]{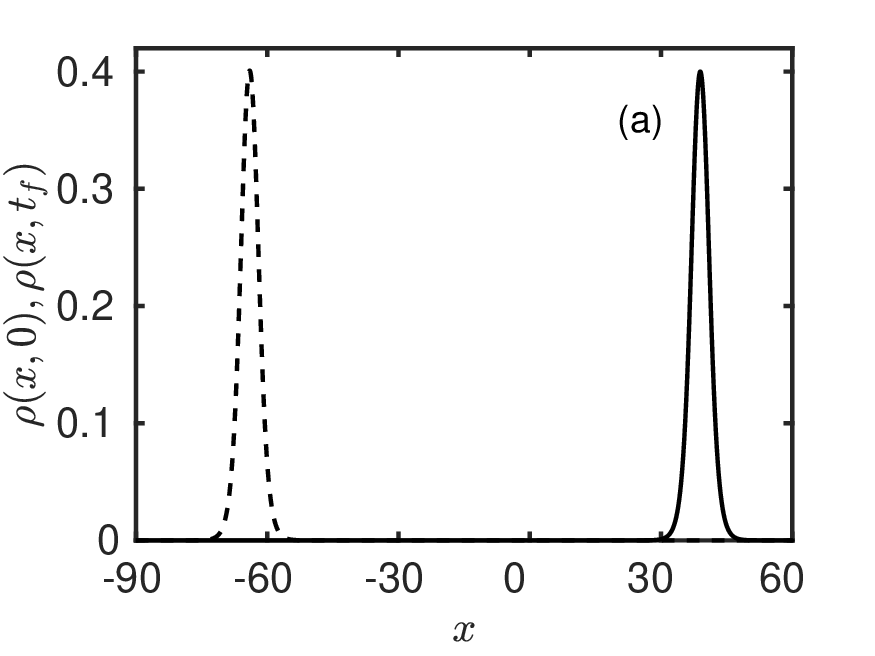} &
		\includegraphics[width=0.4\linewidth]{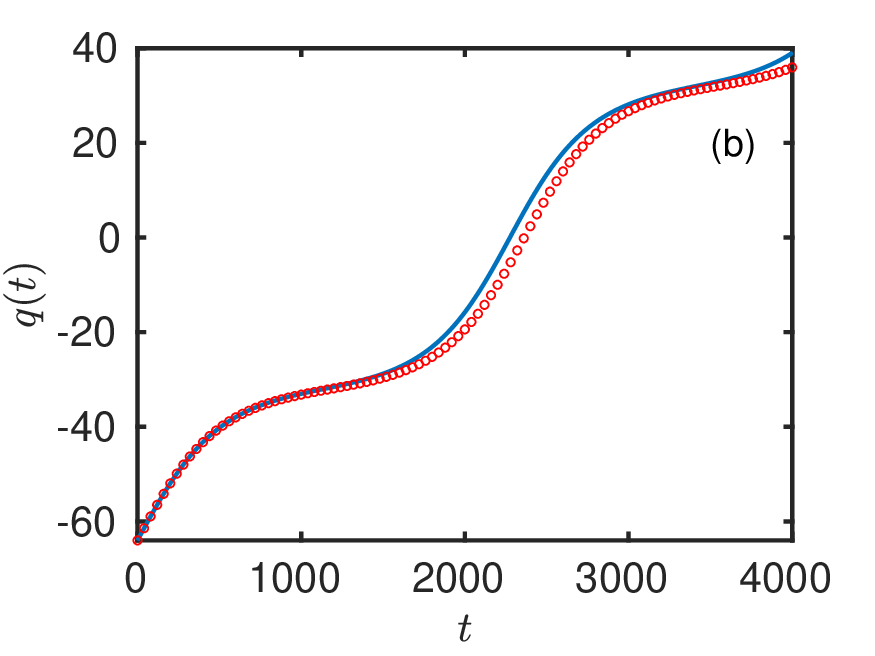} \\
		\includegraphics[width=0.4\linewidth]{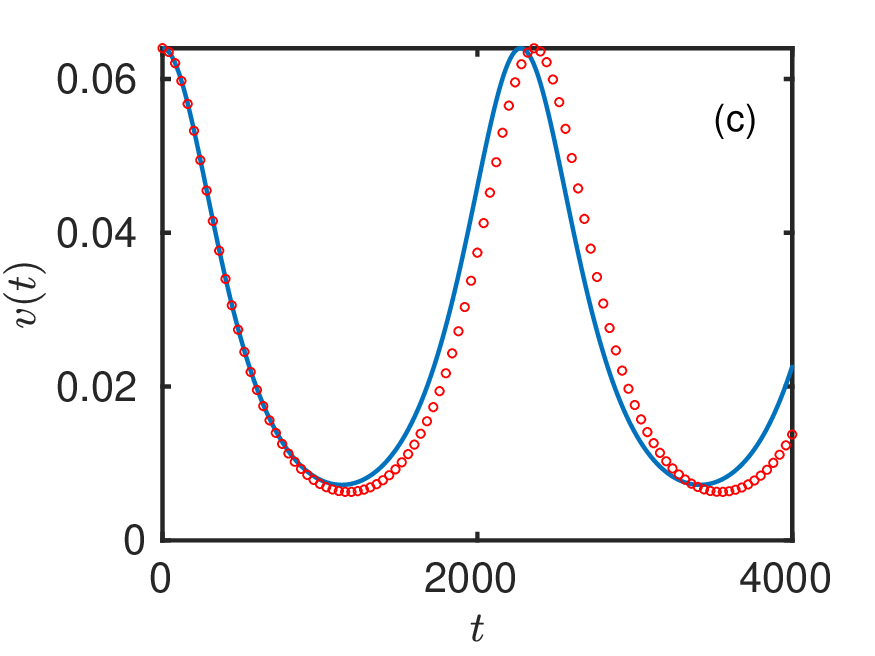} &
		\includegraphics[width=0.4\linewidth]{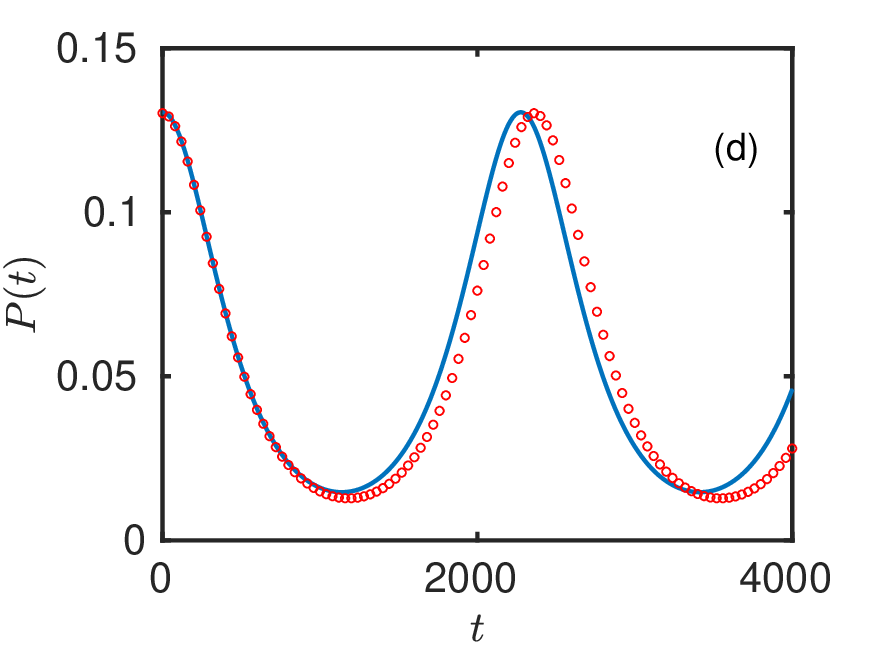} 
	\end{tabular}
	\caption{Soliton dynamics with the periodic potential $V(x)=-\epsilon\,\cos(k\,x)$, for $\omega=0.9$, $\epsilon=0.001$, and $k=\pi/32$.  Panel (a): soliton profiles at $t=0$ (dashed line)  and at $t_f=4000$ (solid line).  
		Panels (b), (c) and (d): comparison of the simulation results for the soliton position $q(t)$, velocity $v(t)$, and momentum $P(t)$ (solid lines) with the CC results (red circles).
		Initial conditions: 
		$q(0)=-64$, $v(0)=0.064> v_c\simeq 0.06369$.}
	\label{fig8}
\end{figure}

From Eq. (\ref{eq36}):
\begin{equation}
\label{eq85}
U(q,\dot{q})=-\epsilon\,\int_{-\infty}^{+\infty}dz\,\cos\left[k\,\left(\frac{z}{\gamma}+q\right)
\right]\,\frac{\rho(z)}{\gamma}.
\end{equation}
In the nonrelativistic regime $\dot{q}^2 \ll 1$ and 
\begin{equation}
\label{eq86}
U(q)=-\epsilon\,\cos(k\,q)\,I_4,
\end{equation}
with 
\begin{equation}
\label{eq87}
I_4=\int_{-\infty}^{+\infty}dz\,\cos(k\,z)\,\frac{\rho(z)}{\gamma}.
\end{equation}
The equation of motion (\ref{eq45}) is 
\begin{equation}
\label{eq88}
\frac{d}{dt}(M_0\,\dot{q}) =-\frac{\partial U}{\partial q},
\end{equation}
which yields the pendulum equation 
\begin{equation}
\label{eq89}
\ddot{q}+C\,\sin(k\,q) =0,
\end{equation}
where $C=\frac{\epsilon\,k\,I_4}{M_0}$. 
Choosing  $\dot{q}(0)=v_0$ we obtain the solution 
\begin{equation}
\label{eq90}
q(t)=q(0)+\frac{2}{k}\,am\left[\frac{k\,v_0}{2}\,t,m\right],
\end{equation}
where $q(0)$ is the initial position and 
$m=4\,C/(k\,v_0^2)$ is the modulus in the Jacobi amplitude $am$. 
At $m=1$, there is a critical velocity
\begin{equation}
\label{eq91}
v_c=\sqrt{\frac{4\,\epsilon\,I_4}{M_0}}.
\end{equation}
For $v_0<v_c$ periodic motion is obtained because the soliton remains confined in a potential well, while for $v_0>v_c$ the motion is unbounded. We have explored numerically three cases: $v_0 \ll v_c \ll 1$ (Fig.\ \ref{fig6}), $v_0$ slightly below $v_c$ (Fig.\ \ref{fig7}), and $v_0$ slightly above $v_c$ (Fig.\ \ref{fig8}). In all cases, the agreement of the CC theory with the simulations is very good. 

\begin{figure}[ht!]
	\centering
	\begin{tabular}{cc}
		\includegraphics[width=0.4\linewidth]{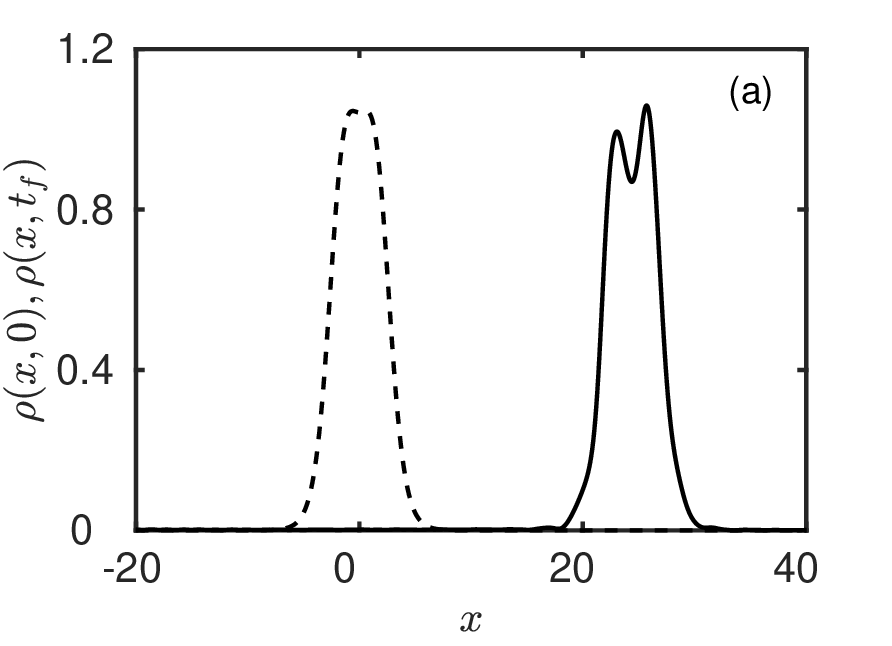} &
		\includegraphics[width=0.4\linewidth]{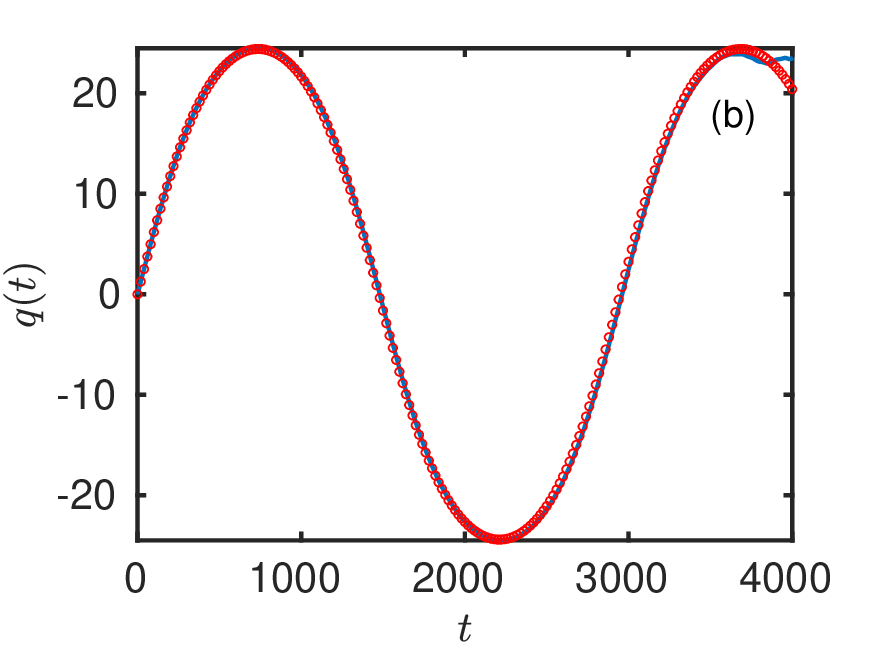} \\
		\includegraphics[width=0.4\linewidth]{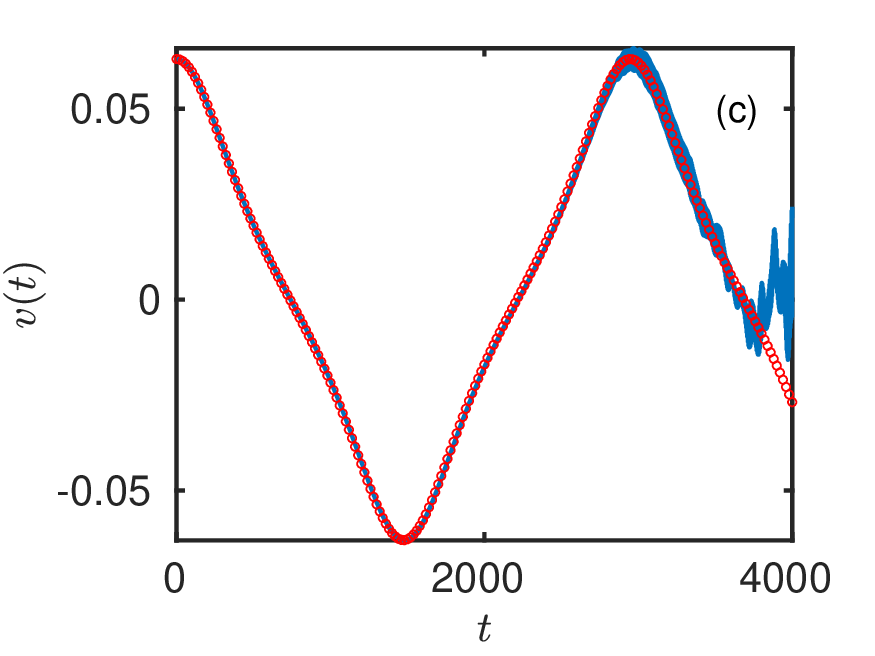} &
		\includegraphics[width=0.4\linewidth]{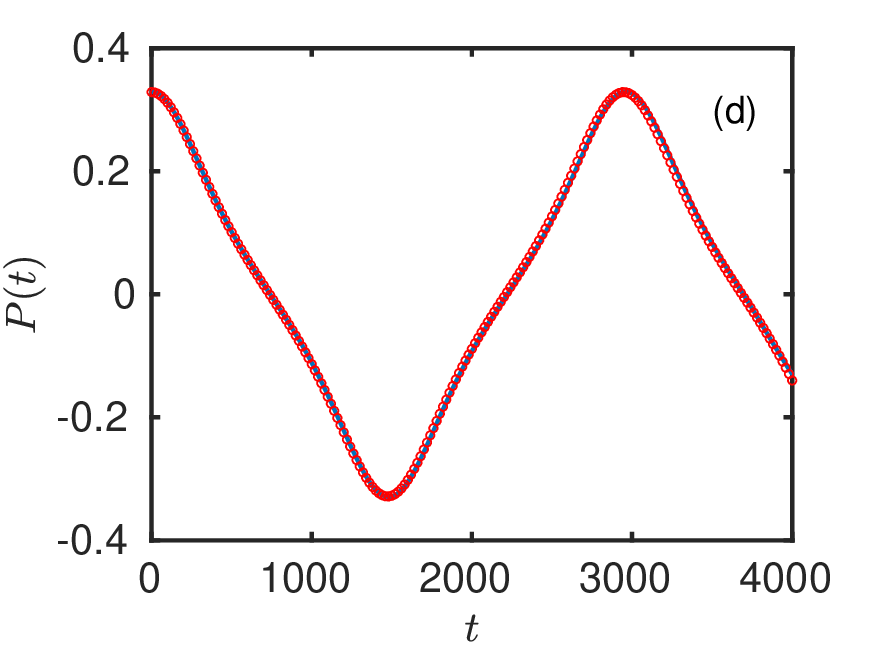} 
	\end{tabular}		
	\caption{Soliton dynamics with the periodic potential $V(x)=-\epsilon\,\cos(k\,x)$, for $\omega=0.74$, $\epsilon=0.001$, and $k=\pi/32$.  Panel (a): soliton profiles at $t=0$ (dashed line)  and at $t_f=4000$ (solid line).  
		Panels (b), (c) and (d): comparison of the simulation results for the soliton position $q(t)$, velocity $v(t)$, and momentum $P(t)$ (solid lines) with the CC results (red circles).
		Initial conditions:     
		$q(0)=0$, $v(0)=0.063< v_c\simeq 0.06369$. }
	\label{fig7b}
\end{figure}

 In Fig.\ \ref{fig7b} we have plotted the case with $v_0$ slightly below $v_c$ and  $\omega=0.74$, close to the upper end of the lower interval 
$1/\sqrt{2}<\omega<3/4$.  The lifetime of the soliton is rather long but eventually becomes unstable. Again, the closer to the lower end of the interval  the frequency is, the shorter the soliton lifetime is. 

\section{Summary}
  
We have investigated the dynamics of solitons in a ABS nonlinear spinor model,
where the solitons are subjected to several potentials. The linear terms in this
model are the same as the linear terms in the  Gross-Neveu equation, but the nonlinear
terms of the two models differ. We have developed a Collective Coordinates
Theory for the  perturbed  ABS model which yields a Lagrangian  that has the same expression 
(apart from the coefficients and an irrelevant phase) as 
the Lagrangian of the Collective Coordinates Theory for the  Gross-Neveu equation. 
 Although, there is a part in the Lagrangian coming from the nonlinear term, it can be rewritten as a linear combination of the mass and charge. 

Thus, the dynamics of the solitons in the two models are \textit{qualitatively} the same, but not
\textit{quantitatively},   due to the differences in the  particle masses and in the amplitudes of the effective potential. The mass (charge) related to the ABS model is always greater (less) than the mass (charge) of the Gross-Neveu model. These differences, however, cause solely slight deviations between the soliton positions of the two models for the considered potentials and chosen parameters. This is very remarkable because  the dynamics of solitons
in models with different nonlinearities are normally qualitatively \textit{different}.

The time evolution of the soliton position is given by a Newtonian-like equation of
motion for a relativistic mass, driven by an effective force. This force is defined by
derivatives of an effective potential with respect to the soliton position and
velocity. This potential is defined as an integral which contains the charge density
of the soliton and the potential in which the soliton moves. The Newtonian-like
equation of motion is solved analytically or numerically. The results for the time
evolution of the soliton position are confirmed by simulations, that is, by numerical
solutions of the  ABS spinor model with a potential, taking as initial condition the
exact soliton solution of the model without a potential. Specifically, when the ramp potential is considered, the soliton is accelerated.
When the initial soliton is in a harmonic potential, its center of mass oscillates
around the minimum of the potential. Finally, if the potential is periodic,  the soliton can be trapped in a well or displays unbounded motion, depending on its initial velocity.

The shape and the lifetime of the solitons depend on $\omega$, which is the only
parameter in the exact soliton solution. For $0.75< \omega <1$, the charge density of
the soliton is bell-shaped and the soliton is stable, i.e. it has an infinite lifetime.
This holds for the three potentials which were considered and it also holds when
the soliton velocity is in the relativistic regime. For $1/\sqrt{2}< \omega <0.75$, the
charge density has two humps and the soliton is unstable.When $\omega$ is
situated close to the upper end of the interval, the lifetime is long. Close to the
lower end  the lifetime becomes shorter.

\section{Appendix. \label{apA}}

In the variational approach, the components of the spinor field  are approximated by  trial functions, $u(x,\{Y_{i}\})$ and $v(x,\{Y_{i}\})$, ($i=1,\cdots,N$) of  $N$ collective coordinate ${Y_i}$. 
These variables satisfy equations of motion, which are obtained from the Lagrange equations 
\begin{eqnarray} \label{eq-LCC}
\frac{d}{dt}\,\frac{\partial L}{\partial \dot{Y}_{i}} - \frac{\partial L}{\partial Y_{i}}=0,
\end{eqnarray}
or equivalently
\begin{eqnarray} \label{eq-LCC1}
\int_{-\infty}^{+\infty} dx \, \left[\frac{d}{dt}\,\frac{\partial {\cal L}}{\partial \dot{Y}_{i}} - 
\frac{\partial {\cal L}}{\partial Y_{i}} \right]=0,
\end{eqnarray}
 where the lagrangian density ${\cal L}={\cal L}(u,u^{*},u_{x},u^{*}_{x},u_{t},u^{*}_{t},v,v^{*},v_{x},v^{*}_{x},v_{t},v^{*}_{t})$ Taking into account that
 \begin{eqnarray} 
\frac{d}{dt}\,\frac{\partial {\cal L}}{\partial \dot{Y}_{i}} & =&
\frac{d}{dt}\, \left(\frac{\partial {\cal L}}{\partial u_{t}} \frac{\partial u}{\partial Y_{i}} \right)+ 
\frac{d}{dt}\, \left(\frac{\partial {\cal L}}{\partial u^{*}_{t}} \frac{\partial u^{*}}{\partial Y_{i}} \right)+
\frac{d}{dt}\, \left(\frac{\partial {\cal L}}{\partial v_{t}} \frac{\partial v}{\partial Y_{i}} \right)+ 
\frac{d}{dt}\, \left(\frac{\partial {\cal L}}{\partial v^{*}_{t}} \frac{\partial v^{*}}{\partial Y_{i}} \right), \nonumber 
\\
\nonumber
\frac{\partial {\cal L}}{\partial Y_{i}} & =& 
\frac{\partial {\cal L}}{\partial u} \frac{\partial u}  {\partial Y_{i}} + 
\frac{\partial {\cal L}}{\partial u^{*}} \frac{\partial u^{*}}  {\partial Y_{i}} +
\frac{\partial {\cal L}}{\partial u_{x}} \frac{\partial u_{x}}  {\partial Y_{i}}+
\frac{\partial {\cal L}}{\partial u^{*}_{x}} \frac{\partial u^{*}_{x}}  {\partial Y_{i}}+
\frac{\partial {\cal L}}{\partial u_{t}} \frac{\partial u_{t}}  {\partial Y_{i}}+
\frac{\partial {\cal L}}{\partial u^{*}_{t}} \frac{\partial u^{*}_{t}}  {\partial Y_{i}}+ \\
\nonumber
&&
\frac{\partial {\cal L}}{\partial v} \frac{\partial v}  {\partial Y_{i}} + 
\frac{\partial {\cal L}}{\partial v^{*}} \frac{\partial v^{*}}  {\partial Y_{i}} +
\frac{\partial {\cal L}}{\partial v_{x}} \frac{\partial v_{x}}  {\partial Y_{i}}+
\frac{\partial {\cal L}}{\partial v^{*}_{x}} \frac{\partial v^{*}_{x}}  {\partial Y_{i}}+
\frac{\partial {\cal L}}{\partial v_{t}} \frac{\partial v_{t}}  {\partial Y_{i}}+
\frac{\partial {\cal L}}{\partial v^{*}_{t}} \frac{\partial v^{*}_{t}}  {\partial Y_{i}}, 
\end{eqnarray} 
then Eq.\ (\ref{eq-LCC1}) becomes
\begin{eqnarray}\nonumber
\int_{-\infty}^{+\infty} dx \left\{ \frac{d }{dt} \left(\frac{\partial {\cal L}}{\partial u_{t}}\right) + 
\frac{d }{dx} \left(\frac{\partial {\cal L}}{\partial u_{x}}\right)-
\frac{\partial {\cal L}}{\partial u}\right\} \frac{\partial u}{\partial Y_{i}}  + \\ \nonumber 
\left\{  
\frac{d }{dt} \left(\frac{\partial {\cal L}}{\partial u^{*}_{t}}\right) + 
\frac{d}{dx} \left(\frac{\partial {\cal L}}{\partial u^{*}_{x}}\right)-
\frac{\partial {\cal L}}{\partial u^{*}} \right \} \frac{\partial u^{*}}{\partial Y_{i}} + \\ \nonumber 
\left\{ \frac{d }{dt} \left(\frac{\partial {\cal L}}{\partial v_{t}}\right) + 
\frac{d }{dx} \left(\frac{\partial {\cal L}}{\partial v_{x}}\right)-
\frac{\partial {\cal L}}{\partial v}\right\} \frac{\partial v}{\partial Y_{i}}  + \\ \nonumber 
\left\{  
\frac{d }{dt} \left(\frac{\partial {\cal L}}{\partial v^{*}_{t}}\right) + 
\frac{d}{dx} \left(\frac{\partial {\cal L}}{\partial v^{*}_{x}}\right)-
\frac{\partial {\cal L}}{\partial v^{*}} \right \} \frac{\partial v^{*}}{\partial Y_{i}}
  = 0, 
\end{eqnarray}
under the assumption
\begin{eqnarray} \nonumber
&&\left[\frac{\partial {\cal L}}{\partial u_{x}} \, \frac{\partial u}{\partial Y_{i}} +\frac{\partial {\cal L}}{\partial u_{x}^{*}} \, \frac{\partial u^{*}}{\partial Y_{i}}+
\frac{\partial {\cal L}}{\partial v_{x}} \, \frac{\partial v}{\partial Y_{i}} +\frac{\partial {\cal L}}{\partial v_{x}^{*}} \, \frac{\partial v^{*}}{\partial Y_{i}}
\right]_{x \to +\infty} - \\ \label{eq-LCC2}
&&\left[\frac{\partial {\cal L}}{\partial u_{x}} \, 
\frac{\partial u}{\partial Y_{i}} +\frac{\partial {\cal L}}{\partial u_{x}^{*}} \, 
\frac{\partial u^{*}}{\partial Y_{i}}+
\frac{\partial {\cal L}}{\partial v_{x}} \, 
\frac{\partial v}{\partial Y_{i}} +\frac{\partial {\cal L}}{\partial v_{x}^{*}} \, 
\frac{\partial v^{*}}{\partial Y_{i}}
\right]_{x \to -\infty} = 0. 
\end{eqnarray}
Therefore, if the assumption (\ref{eq-LCC2}) is satisfied by the ansatz, the variational approach is equivalent to a projection technique,  where one projects the spinor field equation and its complex conjugate equation onto the basis 
$\frac{\partial u^\star}{\partial Y_{i}}$, $\frac{\partial v^\star}{\partial Y_{i}}$, $\frac{\partial u}{\partial Y_{i}}$, $\frac{\partial v}{\partial Y_{i}}$, respectively; and finally one adds it all. 
 This relationship between the variational approach and the projection technique is a  generalization of the proof carried out  
for the perturbed Nonlinear Schr\"odinger equation in \cite{quintero:2010}.

%%%%%%%%%%%%%%%%%%%%
\ack 

The authors thank Igor Barashenkov, Cape Town, South Africa, for discussions on this paper. 
F.G.M. acknowledges financial support and hospitality of the University of Seville.  
N.R.Q. acknowledges the financial support from the Alexander von Humboldt Foundation 
and the hospitality of the Physikalisches Institut at the University of Bayreuth (Germany) during the development of this work. N.R.Q. also acknowledges financial support from the Ministerio de Econom\'{\i}a y Competitividad of Spain through  FIS2017-89349-P.
B.S-R. acknowledges financial support from the Ministerio de Ciencia e Innovación through PGC2018-093998-B-I00.

%%%%%%%%%%%%%%%%%%%%%%% FIG 1 %%%%%%%%%%%%%%%%%%%%%%%%%%%%%%%%%%%%%%%%%%%

%%%%%%%%NEW FIGURE 2 %%%%%%%%%%%%%%%%%%%%%%%%%%%%%%%%%%%

%%%%%%%%%%%%%%%%%%%FIGURE 3%%%%%%%%%%%%%%%%%%%%%%%%%%%%%%%%%%%%%%%

%%%%%%%%%%%%%%%%%%%%%%FIGURE 4%%%%%%%%%%%%%%%%%%%%%%%%%%%%%%%%%%%
%%%%%%%%%%%%%%%%%%%%%%%%%%%%%%%%%%%%%%%%%%%%%%%%%%%%%%%%%%%
%%% FIGURE 5
%%%%%%%%%%%%%%%%%%%%%%%%%%%%%%%%%%%%%%%%%%%%%%%%%%%%%%%%%
%%%%%%%%%%%%%%%%%%%FIGURE 6  %%%%%%%%%%%%%%%%%%%%%%%%%%%%%%%%%%%%%%

%%%%%%%%%%%%%%%%%%%%%%%%FIGURE 7%%%%%%%%%%%%%%%%%%%%%%%%%%%%%%%%%
%%%%%%%%%%%%%%%%%%%%%%FIGURE 8%%%%%%%%%%%%%%%%%%%%%%%%%%%%%%%%%%%

%%%%%%%%%%%%%%%%%%%%%%%FIGURE 9%%%%%%%%%%%%%%%%%%%%%%%%%%%%%%%%%%

%%%%%%%%%%%%%%%%%%%%%%%%FIGURE 10%%%%%%%%%%%%%%%%%%%%%%%%%%%%%%%%

\section*{ORCID iDs}

Franz G Mertens https://orcid.org/0000-0002-0574-2279 \\
Bernardo Sánchez-Rey https://orcid.org/0000-0003-3170-154X\\
Niurka R Quintero https://orcid.org/0000-0003-3503-3040

\end{document}